\documentclass[11pt, aps,prl,amsmath,nofootinbib]{revtex4}
\pdfoutput=1
\usepackage{amsmath,amssymb,amsfonts,dcolumn,color,graphicx,graphics,latexsym,placeins,epsfig,braket,textcomp,hyperref,bm,url}
\usepackage[utf8]{inputenc}

\def\R{\mathcal{R}}
\def\refc{Ref.~\cite}
\def\theory{the GCG model}
\def\theoryj{the GCJG model}
\def\binv{\frac{\rho_i^{1+\alpha}}{A}}
\def\binvtext{\rho_i^{1+\alpha}/A}
\def\fn{\mathcal{G}}

\newcommand{\be}{\begin{equation}}
\newcommand{\ee}{\end{equation}}
\newcommand{\bea}{\begin{eqnarray}}
\newcommand{\eea}{\end{eqnarray}}
\newcommand{\eqn}[1]{(\ref{#1})}
\newcommand{\eq}[1]{Eq.~(\ref{#1})}
\newcommand{\eqs}[1]{Eqs.~(\ref{#1})}
\newcommand{\figr}[1]{Fig.~\ref{#1}}

\newcommand{\aint}{$-1 < \alpha \le 1$}
\newcommand{\ainttwo}{$0 < \alpha < 1$}
\newcommand{\aintthree}{$\alpha <-1$}
\newcommand{\anewint}{$\alpha <1$}
\newcommand{\aconstraint}{$\alpha \le -5/3$}
\newcommand{\aconstraintxi}{$\alpha < -1.6$}

\newcommand{\mpl}{m_{\rm Pl}}
\newcommand{\eh}{\epsilon_H}
\newcommand{\etah}{\eta_H}
\newcommand{\arcsinh}{\text{arcsinh}}
\newcommand{\xh}{\xi_H}

\newcommand{\sech}{\textrm{sech}}

\newcommand{\sd}{\textrm{sd}}

\begin{document}

\title{No slow-roll inflation à la Generalized Chaplygin Gas in General Relativity}
\author{Alexander Gallego Cadavid, J.R. Villanueva}
% \email{alexander.gallego@uv.cl}
% \email{jose.villanueva@uv.cl}
\affiliation{
Instituto de Física y Astronomía, Universidad de Valparaíso,  Avenida Gran Bretaña 1111, Playa Ancha, Valparaíso, Chile
}

\begin{abstract}
The Generalized Chaplygin Gas (GCG) model is characterized by the equation of state $P = -A \rho^{-\alpha}$, where $A>0$ and $\alpha < 1$. The model has been extensively studied due to its interesting properties and applicability in several contexts, from late-time acceleration to primordial inflation. Nonetheless we show that the inflationary slow-roll regime cannot be satisfied by most of the parameter space of the GCG model when General Relativity (GR) is considered. In particular, although the model has been applied to inflation with $0 < \alpha < 1$, we show that for $-1 < \alpha \le 1$ there is no expansion of the Universe but an accelerated contraction. For $\alpha \le -5/3$, the second slow-roll parameter $\eta_H$ is larger than unity, so there is no sustained period of inflation. Only for $\alpha$ very close to -1 the model produces enough $e$-folds, thus greatly reducing its parameter space. Moreover, we show that the model is ruled out by the Planck 2018 results. Finally, we extend our analysis to the Generalized Chaplygin-Jacobi Gas (GCJG) model. We find that the introduction of a new parameter does not change the previous results. We thus conclude that the violation of the slow-roll conditions is a generic feature of the GCG and GCJG models during inflation when GR is considered and that the models are ruled out by the Planck 2018 results.

\end{abstract}

\maketitle

\section{Introduction}
An important prediction of inflationary cosmology is that there should be small departures from the large-scale
homogeneity observed in the Universe~\cite{Akrami:2018vks, Akrami:2018odb}. Inflation predicts that these perturbations have a characteristic spectrum. During inflation, space-time itself fluctuates quantum mechanically about a background in an accelerated expanding phase~\cite{Akrami:2018vks, Akrami:2018odb, Guth:1980zm, Linde:1981mu, Starobinsky:1998mj, Martin:2013tda, Wang:2013eqj,  Castiblanco:2019mgb, xc}. These quantum fluctuations are the seeds for the observed structures and CMB anisotropies since the microscopic fluctuations were spread out to macroscopic scales, where they eventually become classical fluctuations in space-time~\cite{Akrami:2018vks, Akrami:2018odb, Martin:2013tda, Wang:2013eqj,  Castiblanco:2019mgb, xc}.

The simplest models of cosmic inflation involve a scalar field $\phi$, called the inflaton, slowly rolling down a very flat energy potential $V(\phi)$~\cite{Akrami:2018vks, Akrami:2018odb,  Guth:1980zm, Linde:1981mu, Starobinsky:1998mj, Martin:2013tda, Wang:2013eqj, Castiblanco:2019mgb, xc}. During inflation most of the Universe's energy density is in the form of $V(\phi)$. In these circumstances the field behaves like a fluid with negative pressure and thus powers an almost exponential cosmic expansion. Inflation ends when the inflaton's kinetic energy is larger than its potential energy, which occurs in the steeper part of $V(\phi)$. Nonetheless, before the end of inflation, it is crucial to have enough slow rolling for the scalar field in order to solve the horizon and flatness problems with inflation~\cite{Akrami:2018vks, Akrami:2018odb,  Martin:2013tda, Wang:2013eqj, Castiblanco:2019mgb, xc}. This corresponds to a growing of the scale factor $a$ of the order of $e^{60}$~\cite{Akrami:2018vks, Akrami:2018odb,  Guth:1980zm, Linde:1981mu, Starobinsky:1998mj, Martin:2013tda, Wang:2013eqj, Castiblanco:2019mgb, xc, Chakraborty:2018scm}. 

Recently, the generalized Chaplygin gas (GCG) model has been discussed widely in cosmological contexts. The model is characterized by a fluid with an exotic equation of state
\be \label{eosch} P = -\frac{A}{\rho^{\alpha}} , 
\ee
where $A>0$ and  $\alpha < 1$ \cite{Bento:2002ps, Sen:2005sk, Dinda:2014zta, Chimento:2005au}. The model corresponds to a generalized Nambu-Goto action which can be interpreted as a perturbed $d$-brane in a $(d+1,1)$ spacetime \cite{Bento:2002ps}. The case $\alpha=1$ reproduces the pure Chaplygin gas model in Ref.~\cite{Kamenshchik:2001cp}, while $\alpha =-1$ corresponds to a cosmological constant \cite{Kamenshchik:2001cp, Bento:2002ps, Sen:2005sk, Dinda:2014zta, Chimento:2005au, 2009EPJC...63..349L, Fabris:2001tm, Makler:2002jv, Bilic:2001cg, Dev:2002qa}. In the late-time cosmological context and for $\alpha <-1$, \eq{eosch} leads to three new versions of the GCG model: an early phantom model, a late phantom model, and a transient model \cite{Sen:2005sk}. 

Originally, the Chaplygin gas model was introduced to explain late-time acceleration without dark energy in the context of General Relativity (GR)~\cite{Kamenshchik:2001cp, Bento:2002ps, Sen:2005sk, Dinda:2014zta, Chimento:2005au, 2009EPJC...63..349L, Fabris:2001tm, Makler:2002jv, Bilic:2001cg, Dev:2002qa}. So far there have been different modifications to the GCG proposal such as modified Chaplygin gas in brane-world~\cite{Herrera:2008us, Mak:2005iq, Jawad:2016key} and GCG in modified gravity  \cite{Chimento:2005au, Barreiro:2004bd, Bertolami:2006zg}. 

A first approach to single-field inflation using the GCG as the inflaton field was done in Ref.~\cite{delCampo:2013uta} using \ainttwo, where the value $\alpha =  0.2578 \pm 0.0009$ was found from the Planck 2013 data. Later the GCG inflation was studied in light of Planck where it was shown that the GCG is not a suitable candidate for inflation in the context of GR for $\alpha <-1$ \cite{Dinda:2014zta}. In this work the authors found that, in order to obtain the observational Planck+ WP and BICEP2 bounds for the spectral index and the tensor-to-scalar ratio, the number of $e$-folds before the end of inflation at horizon exit $N_*$ for the modes would have to be of order $N_* \approx 217$, which is way out of the theoretical bound $50 < N_* < 60$ \cite{Akrami:2018vks, Akrami:2018odb}. More recently, a further generalization of the GCG model was studied using elliptic functions to describe the inflationary epoch \cite{Villanueva:2015ypa, Villanueva:2015xpa}. There the inflaton field was characterized by an equation of state corresponding to a generalized Chaplygin-Jacobi gas (GCJG) 
\be \label{eoschj}
P =-\frac{A\, \kappa}{\rho^{\alpha}} -2\,(1-\kappa)\,\rho +\frac{(1-\kappa)}{A}\,\rho^{2+\alpha}, 
\ee
where $\alpha$ is the GCG parameter, $0 \leq \alpha \leq 1$, and $0\leq \kappa \leq 1$ is the modulus of the elliptic function \cite{Villanueva:2015ypa, Villanueva:2015xpa}. 

In this paper we show that the GCG and GCJG models applied to inflation in Refs.~\cite{delCampo:2013uta} and \cite{Villanueva:2015ypa, Villanueva:2015xpa}, in \eqs{eosch} and \eqn{eoschj} respectively, do not produce an inflationary epoch for the values of the parameters studied there. We then extend the $\alpha$-parameter space to $\alpha<0$ for the models and show that there is a sustained inflationary period only for $\alpha$ very close to -1. Thus, the parameter space for the models is greatly reduced.

The layout of the paper is the following. In Sections \ref{sfi} and \ref{hjfsfi} we review single-field inflation and the Hamilton-Jacobi formalism, respectively. There we emphasis on the number of $e$-folds during inflation and the slow-roll conditions. In Section \ref{igcg} we apply the Hamilton-Jacobi formalism in order to study primordial inflation using a fluid which presents the properties of the GCG. Then in Section \ref{nigcg} we show that for $\alpha>-1$ the GCG model does not produce an inflationary epoch. In Section \ref{nsrigcg} we show that the GCG model produces a burst of inflation only for $\alpha > -5/3$, since the slow-roll conditions are not satisfied for \aconstraint. We also constrain the model from the Planck 2018 results. In Section \ref{nsrigcjggr} we extend our analysis to the case of the GCJG where we find similar results.  The conclusions are presented in Section \ref{conclusions}. In this paper we will restrict ourselves to the study of cosmic inflation using the GCG and GCJG models in the context of GR. Throughout this work we use natural units ($c=\hbar=k_B=1$). 

\section{Single-field inflation}\label{sfi} 
Cosmic inflation is a period of accelerated expansion in the early Universe where the scale factor $a$ behaves like~\cite{Akrami:2018vks, Akrami:2018odb, Guth:1980zm, Linde:1981mu, Starobinsky:1998mj, Martin:2013tda, Wang:2013eqj,  Castiblanco:2019mgb, xc} 
% Cosmic inflation is a period of the Universe’s evolution during which the scale factor $a$ is increasing and accelerating~\cite{Akrami:2018vks, Akrami:2018odb, Guth:1980zm, Linde:1981mu, Starobinsky:1998mj, Martin:2013tda, Wang:2013eqj,  Castiblanco:2019mgb, xc}
\be \label{icond}
\ddot a >0 .
\ee
In the simplest models, inflation is driven by the canonical inflaton $\phi$, slowly rolling the smooth potential energy $V\left(\phi\right)$. Throughout this paper we consider a Friedmann-Lema\^itre-Robertson-Walker (FLRW) metric in a flat Universe
\begin{equation}
{\rm d}s^2 = -{\rm d}t^2 + a^2\left(t\right) {\rm d} \vec{x}\,^2 = a^2\left(\tau\right) \left[-{\rm d}\tau^2 + {\rm d} \vec{x}\,^2\right],
\end{equation}
where $d\tau \equiv dt/a$ is the conformal time and d$\vec{x}\,^2$ is the metric on a maximally symmetric 3–manifold. We assume that the stress energy of the Universe is dominated by the inflaton $\phi$, such that the Einstein field equations of the background metric are given by~\cite{Martin:2013tda, Castiblanco:2019mgb, xc}
\begin{equation}
H^2 \equiv \left({\dot a \over a}\right)^2 = {8 \pi \over 3 m_{\rm Pl}^2} \left[V\left(\phi\right) + {1 \over 2} \dot\phi^2\right] ,
\label{eqbackgroundequation1}
\end{equation}
and
\begin{equation}
\left({\ddot a \over a}\right) = {8 \pi \over 3 \mpl^2} \left[V\left(\phi\right) - \dot\phi^2\right] ,
\label{eqbackgroundequation2}
\end{equation}
where  $H$ is the Hubble parameter, $\mpl \equiv G^{-1/2}$ is the Planck mass, and dots denote derivatives with respect to cosmic time. 

The equation of motion of the spatially homogeneous scalar field is given by~\cite{Martin:2013tda, Castiblanco:2019mgb, xc}
\begin{equation}
\ddot \phi + 3 H \dot \phi + V'\left(\phi\right) = 0,
\label{eqequationofmotion}
\end{equation}
where primes denote derivatives with respect to $\phi$. 

\subsection{Number of $e$-folds and the slow-roll parameters}\label{nesrp}
The amount of inflation is quantified by the ratio of the scale factor at the final time to its value at some initial time $t_i$. This ratio is normally a large number thus it is customary to take the logarithm to give the number of $e$-folds
\be\label{efolds1}
  N(t) \equiv \ln \left[\frac{a(t)}{a(t_i)}\right] = \int_{a_i}^{a} \frac{d \tilde a}{\tilde a} = \int_{t_i}^{t} H(\tilde t)d \tilde t .
\ee
In order to obtain at least $N \sim 60$ we need to impose that $H$ does not change much within a Hubble time $H^{-1}$, i.e., $dH^{-1}/dt \ll 1$ \citep{ xc}. This requisite is equivalent to the first \emph{slow-roll} condition
\begin{equation}\label{sreps}
  \epsilon \equiv -\frac{\dot H}{H^2} \ll 1.
\end{equation}
The second slow-roll condition is given by the requirement that $\epsilon$ does not change much within a Hubble time 
\begin{equation}\label{sreta}
  \eta \equiv \frac{\dot \epsilon}{\epsilon H} \ll 1.
\end{equation}
The slow-roll approximation applies when these parameters are small in comparison to unity, i.e., $\epsilon, |\eta| \ll 1$. As long as $\epsilon <1$, a successful period of inflation can be realized even when $|\eta| >1$ for a few $e$-folds \cite{Starobinsky:1992ts, Adams:2001vc, Lidsey:1995np}. On the other hand the violation of the second slow-roll condition, i.e. $|\eta| >1$, implies that $\epsilon$ grows very rapidly bringing inflation to a swifter end, and thus yielding a small number of $e$-folds \cite{Lidsey:1995np}. As we will see below, this aspect is crucial in our study of the GCG and GCJG models during inflation since we will prove below that, generically, $|\eta| >1$. Hence the GCG and GCJG models produce a small number of $e$-folds.

\section{The Hamilton-Jacobi formalism of single-field inflation} \label{hjfsfi}

In general, the Hubble parameter $H$ will vary as the inflaton field $\phi$ evolves along the potential energy $V\left(\phi\right)$. In some cases a more convenient approach is to express the Hubble parameter directly as a function of the field $\phi$ instead of as a function of time, i.e., $H = H\left(\phi\right)$. In this section we will follow this path, known as the {\it Hamilton-Jacobi} formalism~\cite{Lidsey:1995np, inf, Dinda:2014zta, Kinney:1997ne}.
%In this section we 

We start by differentiating Eq. (\ref{eqbackgroundequation1}) with respect to time from which we obtain \cite{Dinda:2014zta}
\be
2 H\left(\phi\right) H'\left(\phi\right) \dot\phi = - \left(\frac{8\pi}{\mpl^2} \right) H\left(\phi\right) \dot\phi^2,
\ee
and where we used \eq{eqequationofmotion} to eliminate $\ddot \phi$. Substituting back into the definition of $H$ in Eq. (\ref{eqbackgroundequation1}) results in the system of two first-order equations
\bea
\label{phidot}
\dot\phi = && -\frac{\mpl^2}{4\pi} H'\left(\phi\right), \\
\label{Hprime}
\left[H'\left(\phi\right)\right]^2 - \frac{12 \pi}{\mpl^2} H^2\left(\phi\right) = && - \frac{32 \pi^2}{\mpl^4} V\left(\phi\right). 
\eea
These equations are completely equivalent to the second-order equation of motion in \eq{eqequationofmotion}. The second of these is referred to as the {\it Hamilton-Jacobi} equation~\cite{Lidsey:1995np, inf, Dinda:2014zta, Kinney:1997ne}. It allows us to consider $H(\phi)$ as the fundamental quantity to be specified, instead of the usual potential energy $V(\phi)$. In cases where $H(\phi)$ is known, the Hamilton-Jacobi approach is very useful in obtaining several inflationary quantities. For instance, from \eq{phidot} we may obtain an explicit expression for the inflaton field in terms of the cosmological time $t$~\footnote{This is consistent as long as $t$ is a single-valued function of $\phi$.}. From \eq{Hprime} the inflaton potential is given by
\be V(\phi)= \left(\frac{3 \, \mpl^2}{8\,\pi}\right)\left[H^2(\phi) - \frac{\mpl^2}{12 \,\pi}\left[H'(\phi)\right]^2\right]. \label{VCh}
\ee

Moreover, by multiplying \eq{phidot} by $da/d\phi$ we can also obtain an expression for the scale factor in the form
\be \label{aphi} 
a(\phi) =  a_i \exp\left\{-\frac{4\,\pi}{\mpl^2}\,\int_{\phi_i}^\phi \frac{H(\tilde \phi)}{H'(\tilde \phi)}\,d \tilde \phi \right\}.
\ee
Finally, from this last expression, assuming that we have the scalar field as a function of time, we can obtain the scale factor as a function of cosmological time.

In this formalism the slow-roll parameters are defined as~\cite{Lidsey:1995np, inf, Dinda:2014zta, Kinney:1997ne}
\bea
\label{epsilon}
\epsilon_H (\phi) &&\equiv \frac{m_{{\rm Pl}}^2}{4\pi} \left( \frac{H' (\phi) }{H(\phi)} \right)^2  \,, \\
\label{eta}
\eta_H (\phi) &&\equiv \frac{m_{{\rm Pl}}^2}{4\pi} \frac{H''(\phi)}{H(\phi)} \,, \\
\label{xi}
\xi_H^2 (\phi) &&\equiv \Bigl( \frac{m_{{\rm Pl}}^2}{4\pi} \Bigr)^2 \frac{H'(\phi) H'''(\phi)}{H^2(\phi)} \,,
\eea
where we have introduced a third $\xi_H$ slow-roll parameter which is important in the study of scalar perturbations. 

The inflationary condition in \eq{icond}, $\ddot{a}>0$, is precisely equivalent to the condition $\epsilon_H <1$. In order to see this \eq{eqbackgroundequation2} can be written as 
\begin{equation}\label{icond2}
\left({\ddot a \over a}\right) = H^2 \left(\phi\right) \Bigl[1 - \epsilon_H\left(\phi \right) \Bigr],
\end{equation}
such that inflation ends once $\epsilon_H$ exceeds unity. Note that the conditions leading to a violation of the strong energy condition are uniquely determined by the magnitude of $\epsilon_H$ alone~\cite{Lidsey:1995np}. As we stated before, inflation can still proceed if $|\eta_H |$ or $|\xi_H|$ are much larger than unity, though normally such values would drive a rapid variation of $\epsilon_H$ and bring about a swift end to inflation~\cite{Lidsey:1995np, Kinney:1997ne, inf, Romano:2008rr, Arroja:2011yu, a1, a2, a3, Romano:2014kla, GallegoCadavid:2017pol}.

So far we have reviewed cosmic inflation and the Hamilton-Jacobi formalism. In the next sections we will study the GCG model in the Hamilton-Jacobi approach. In Section \ref{nigcg} we will use this formalism to show that, in contradiction to previous results~\cite{delCampo:2013uta, Villanueva:2015ypa, Villanueva:2015xpa}, inflation does not proceed when $\alpha> -1$ for the \theory. Then in Section \ref{nsrigcg} we prove that the slow-roll condition $|\etah| <1$ only holds for $-5/3 <\alpha <-1$.

\section{Inflation à la Generalized Chaplygin Gas} \label{igcg}
In this section we apply the Hamilton-Jacobi formalism in order to study primordial inflation using a fluid which presents the properties of a GCG with equation of state given by \eq{eosch}. 
% Here we will extend the $\alpha$-parameter from \ainttwo~studied in Ref. \cite{delCampo:2013uta} to the more general case of the GCG model with $\alpha < 1$, i.e. including negative values of $\alpha$.

In this formalism the generating function $H(\phi)$ is given by \cite{Dinda:2014zta, delCampo:2013uta}
\be
H(\phi) =H_0 \cosh^{\frac{1}{1+\alpha}}\left[\sqrt{\frac{6 \pi}{\mpl^2}}(1+\alpha)(\phi - \phi_0)\right], 
\label{Hch} 
\ee
where 
\be \label{H0}
H_0 = H(\phi_0) \equiv \sqrt{\frac{8 \pi}{3\,\mpl^2}}\,A^{\frac{1}{2(1+\alpha)}} \, ,
\ee
and $\phi_0$ is an integration constant given by
\be \label{phi0}
\phi_0 =\phi_i - \frac{1}{(1+\alpha)} \sqrt{\frac{\mpl^2}{6\pi}}\, {\text{arcsinh}}\Biggl[ \sqrt{\frac{\rho_i^{1+\alpha}}{A} - 1} \Biggr]  ,
\ee
where $\rho_i = \rho (\phi_i)$ is the energy density at the beginning of inflation and $\binvtext >1$~\cite{Dinda:2014zta}. The condition $\binvtext >1$ comes from both the null energy condition $\rho + P \ge 0$ and the continuity equation $\dot \rho +3\dot a (\rho +P)/a =0$. In the latter case, there is no solution to the continuity equation when $A=\rho^{1+\alpha}_i$~\cite{Dinda:2014zta}. 

Now using \eq{aphi} we calculate the scale factor
\be \label{ach}
a(\phi) =  a_i \exp\left\{-\frac{2}{3(1+\alpha)}\,\int_{\Phi_i}^\Phi \coth[(1+\alpha) \tilde \Phi ] \, d \bigl((1+\alpha) \tilde \Phi \bigr) \right\} = a_i \Biggl( \frac{\sinh[(1+ \alpha) \Phi]}{\sinh[(1+ \alpha) \Phi_i]} \Biggr)^{\frac{-2}{3(1+\alpha)}},
\ee
where for simplicity we use the dimensionless variable
\be \Phi(\phi) \equiv \sqrt{\frac{6\pi}{\mpl^2}}(\phi - \phi_0),
\ee
such that
\be
\Phi_i(\phi) \equiv \sqrt{\frac{6\pi}{\mpl^2}}(\phi_i - \phi_0) = \frac{1}{1+\alpha} \arcsinh{\Biggl[ \sqrt{\frac{\rho_i^{1+\alpha}}{A} - 1} \Biggr]} .  
\ee
In \refc{Dinda:2014zta} the authors calculated the same expression for $a(\phi)$ from the continuity equation.

The slow-roll parameters in \eqs{epsilon} - \eqn{xi} are written as \cite{delCampo:2013uta}
\bea
\label{epsilonch}
\epsilon_H (\phi) &&= \frac{3}{2} \tanh^2\,\left[(1+\alpha)\Phi\right]  \,, \\
\label{etach}
\eta_H (\phi) &&= \frac{3}{2}\,(1+ \alpha \, {\text{sech}}^2\left[(1+\alpha)\Phi\right] )  \,, \\
\label{xich}
\xi_H^2 (\phi) &&= \frac{9}{8}\,\Bigl( 1-2\alpha -4\,\alpha^2+ \cosh\left[2\,(1+\alpha)\Phi\right] \Bigr)
{\text{sech}}^2\left[(1+\alpha)\Phi\right]\,\tanh^2\left[(1+\alpha)\Phi\right]  \,.
\eea
The condition for the end of inflation, $\epsilon_{_{H}}(\Phi_e) = 1$, yields
\be \label{Phie}
\Phi_e (\alpha)= \frac{1}{1+\alpha}\text{arctanh}\left(\sqrt{\frac{2}{3}}\right) ,
\ee
or equivalently
\be \label{phie}
\phi_e (\alpha,A)= \phi_0 + \sqrt{\frac{\mpl^2}{6\pi}} \frac{1}{(1+\alpha)}\text{arctanh}\left(\sqrt{\frac{2}{3}}\right) .
\ee

Finally, the number of $e$-folds $N$ from the beginning of inflation in \eq{efolds1} is given by
\be \label{efoldsch} 
N(\Phi) =-\frac{2}{3 (1+\alpha)}\,\ln\left\{\frac{\sinh\left[(1+\alpha) \Phi \right]}{\sinh \left[(1+\alpha)\Phi_i \right]}\right\},
\ee
thus that the total number of $e$-folds at the end of inflation $N_e$ is
\be \label{Ne} 
N_e(\alpha, A) = \frac{1}{3 (1+\alpha)}\,\ln\left\{\frac12 \left( \binv -1 \right) \right\} ,
\ee
where we use \eq{phi0}. As can be seen $N_e$ depends on the \theory~parameters $\alpha$ and $A$ in such a way that for either $\alpha \to -1$ or $\binvtext \to 1$, it can be arbitrarily large. We can also see that for $\binvtext \ge 3$ \theory~does not yield an inflationary period regardless of the value of $\alpha <-1$, since in this case $N\le 0$. Hence $A$ is further constrained to $\rho_i^{1+\alpha} > A \ge \rho_i^{1+\alpha}/3$.

Unless stated otherwise, we use the following initial conditions in the plots below. Since from Planck 2018~\cite{Akrami:2018odb} the upper bound on the Hubble parameter during inflation is $H_* < 5.4\times 10^{-6} \mpl$, we use for the initial energy density\footnote{See Refs. \cite{Romano:2008rr,  Arroja:2011yu, a1, a2, a3, Romano:2014kla, GallegoCadavid:2017pol} for similar energy scales.} $\rho_i = 1.082 \times 10^{-14} \mpl^4$. For the rest of the parameters we use $a_i=1$, $\phi_i=10 \mpl$, $\alpha <1$, and $\binvtext=1.00001$, where we use $\binvtext$ instead of $A$  for simplicity.

\begin{figure}
 \begin{minipage}{.45\textwidth}
  \includegraphics[scale=0.56]{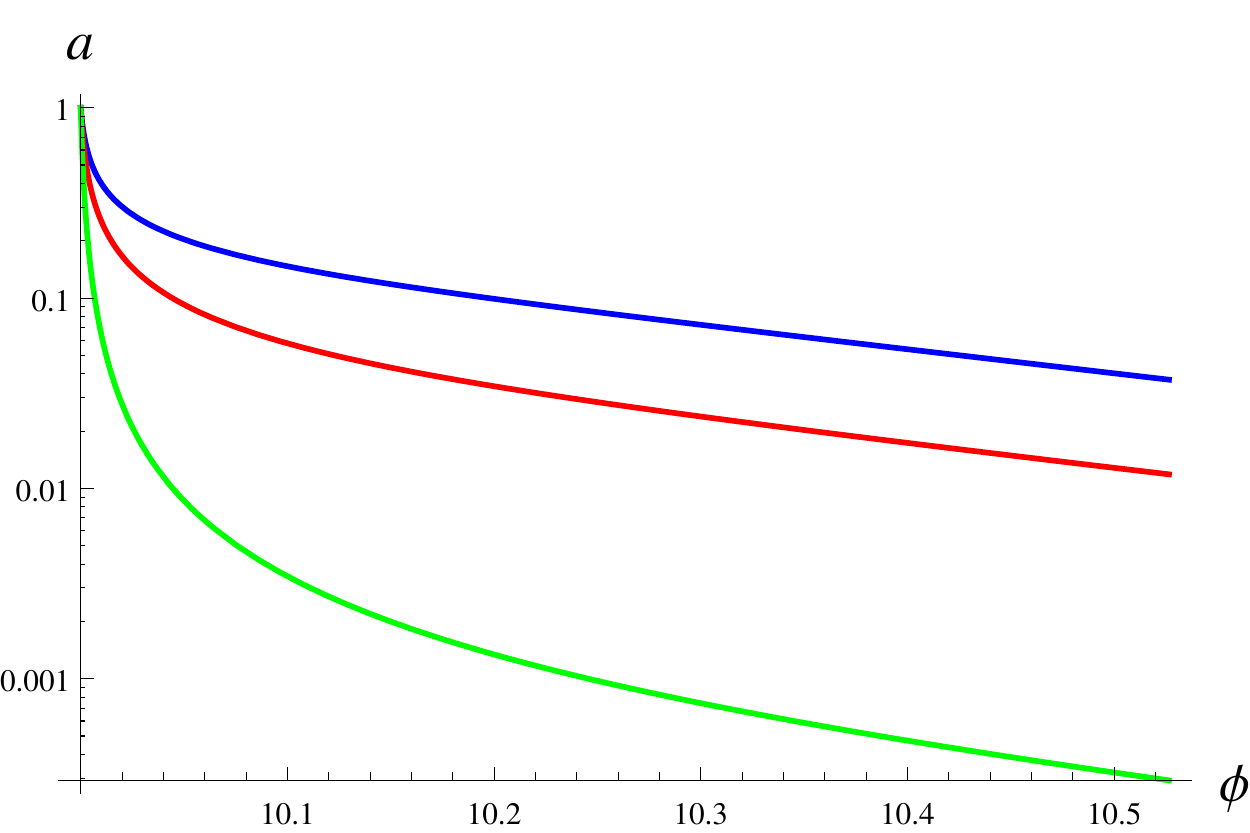}	
  \end{minipage}
 \begin{minipage}{.45\textwidth}
  \includegraphics[scale=0.56]{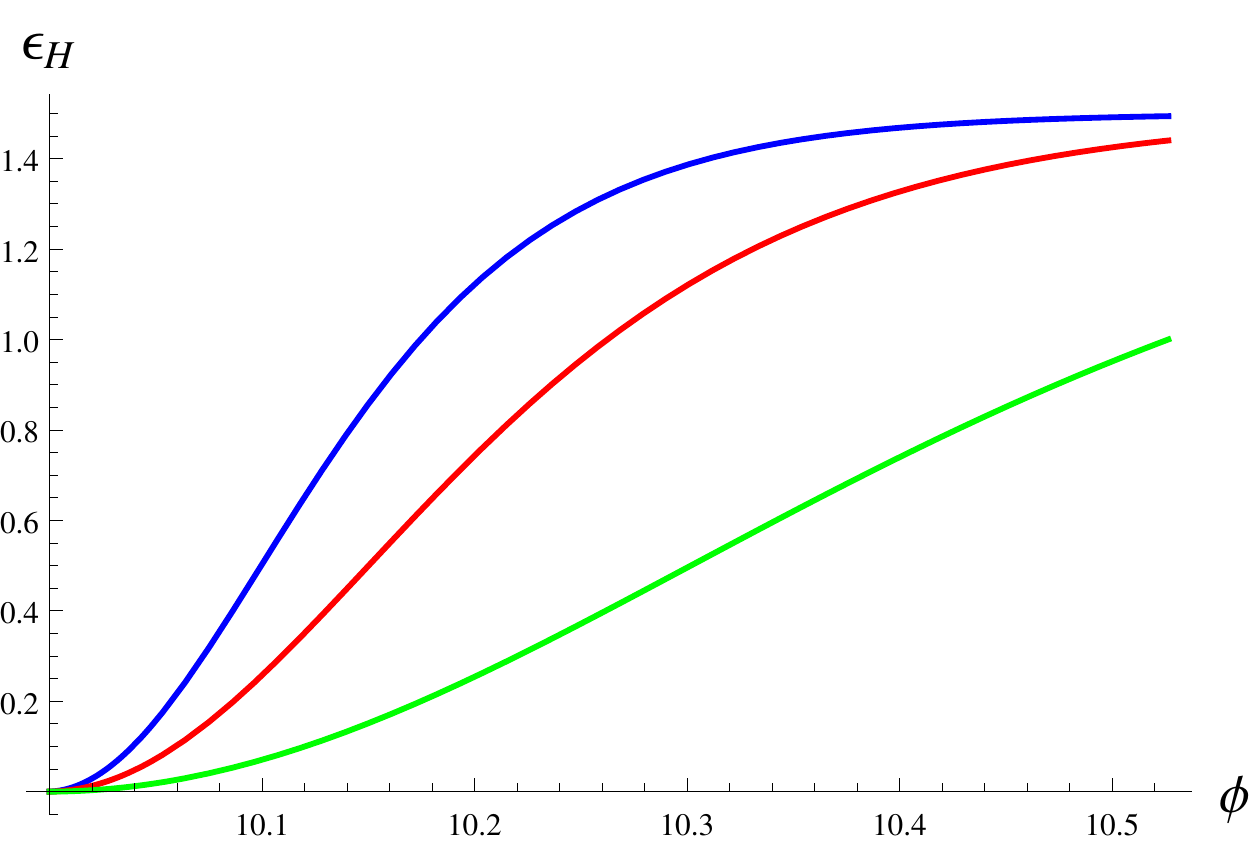}
 \end{minipage}
  \caption{The scale factor $a$ and the slow-roll parameter $\epsilon_H$ are plotted for $\alpha=0.5, 0, -0.5$ and $\binvtext=1.00001$ in the blue, red, and green lines, respectively.  The other values chosen for the parameters are described at the end of Section~\ref{igcg}. Notice that, although $\eh<1$, the scale factor is in fact decreasing. This can also be seen in \eq{ach} where the scale factor is a decreasing function for $\alpha >-1$}
\label{fig:aepsni}
\end{figure}

\section{No inflation à la Generalized Chaplygin Gas ($\alpha >-1$)} \label{nigcg}
In \refc{delCampo:2013uta} the authors studied \theory~in the context of inflation using the interval \ainttwo. In this section we extend the study to $-1< \alpha < 1$ and show that \theory~does not produce an accelerated expansion but an accelerated contraction. 

From \eq{ach} we can see that in fact the scale factor is a decreasing function of $\phi$ for $\alpha >-1$. This equation should be compared with Eq.~(3.7) in Ref.~\cite{delCampo:2013uta}, where the minus sign in the exponent is missing. This was probably the reason why the authors considered \theory~for inflation in the interval \ainttwo. Moreover, in \refc{delCampo:2013uta} only the deceleration parameter $q =-\ddot a a/ \dot a^2$, or equivalently $\eh$, was plotted. Unfortunately this parameter alone does not give information of weather $a$ is increasing or decreasing. 

In \figr{fig:aepsni} the scale factor $a$ and the slow-roll parameter $\eh$ are plotted as functions of $\phi$. There it can be seen that $a$ decreases when $\alpha>-1$ although $\eh<1$. This means that in \theory~the Universe is contracting in an accelerated fashion when $\alpha>-1$.

\begin{figure}
 \begin{minipage}{.45\textwidth}
  \includegraphics[scale=0.56]{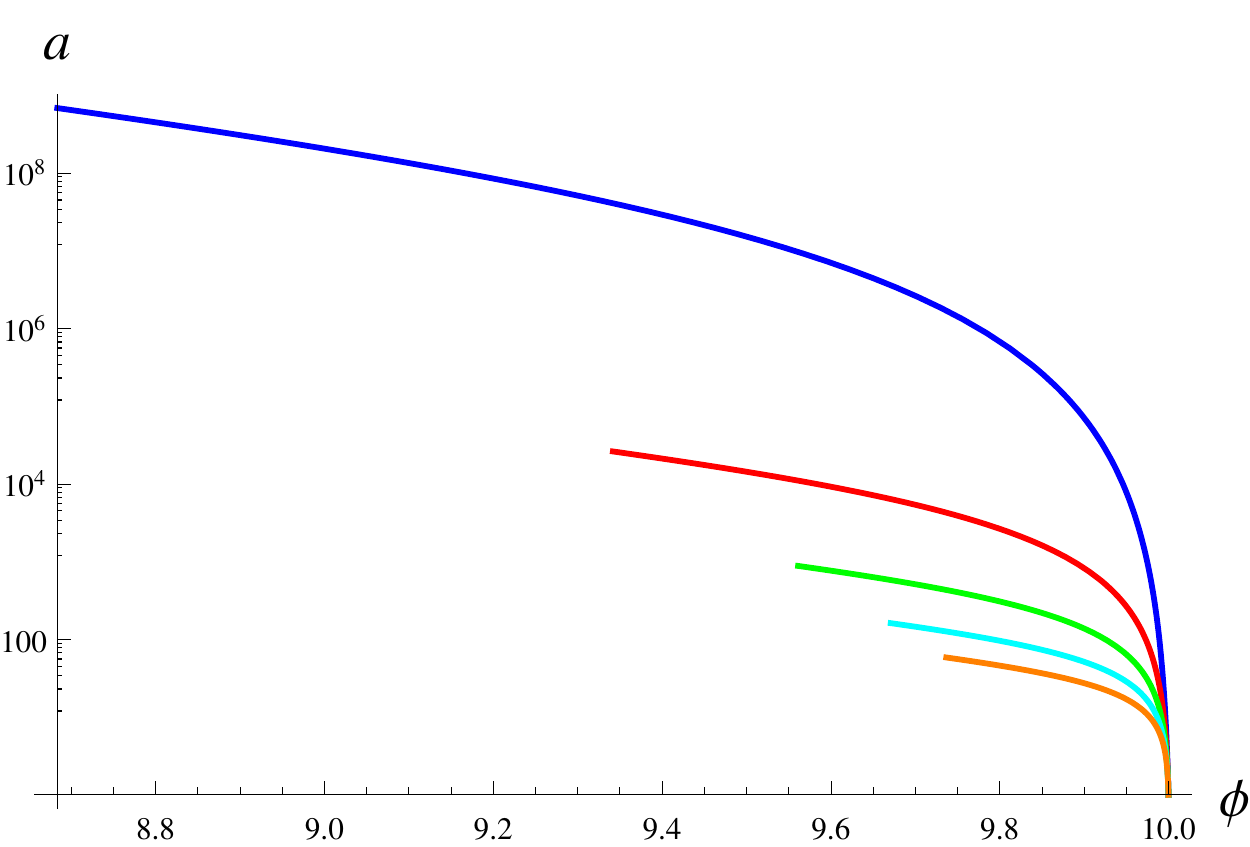}	
  \end{minipage}
 \begin{minipage}{.45\textwidth}
  \includegraphics[scale=0.56]{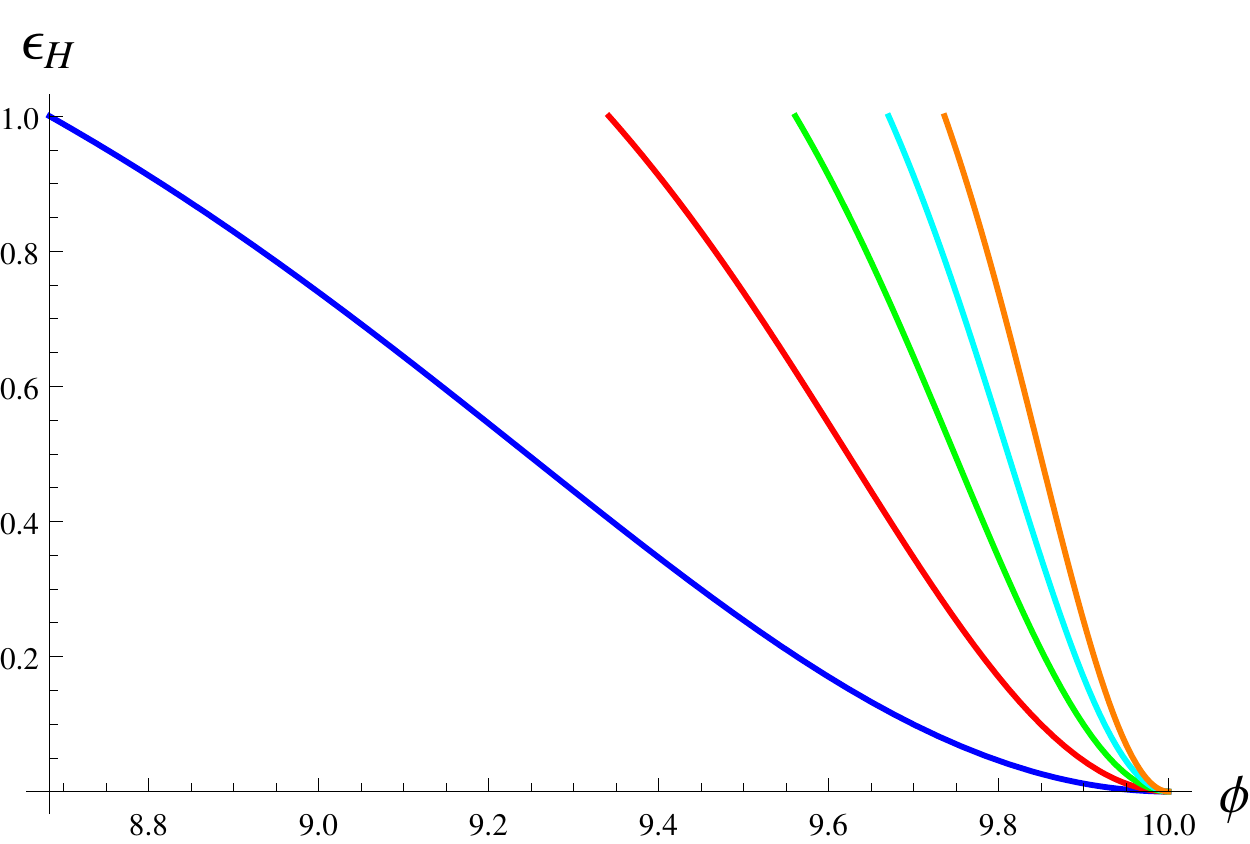}
 \end{minipage}
 \begin{minipage}{.45\textwidth}
  \includegraphics[scale=0.56]{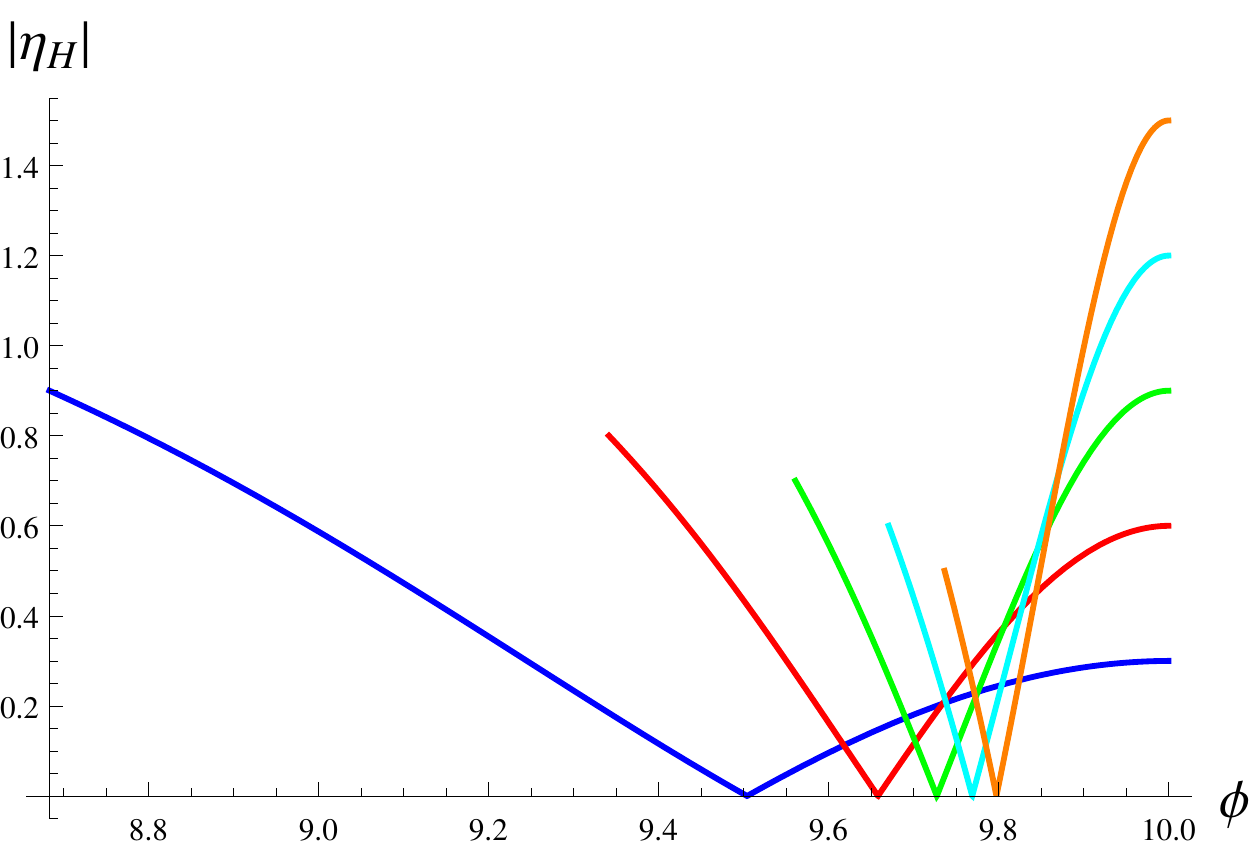}
  \end{minipage}
 \begin{minipage}{.45\textwidth}
  \includegraphics[scale=0.56]{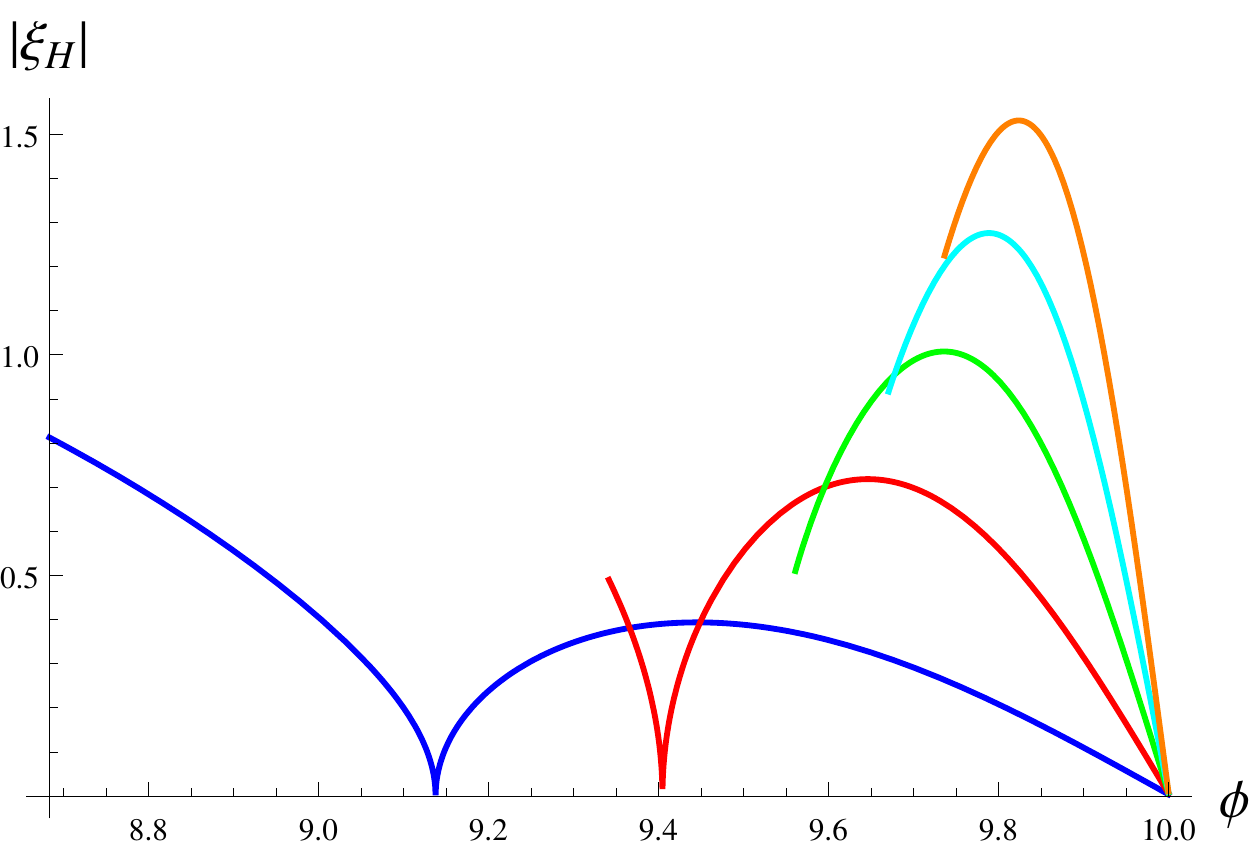}
 \end{minipage}
  \caption{The scale factor $a$ in \eq{ach} and the slow-roll parameters $\epsilon_H ,\eta_H$ and $\xi_H^2$ in \eqs{epsilonch} - \eqn{xich} are plotted for several values of $\alpha$ and up to the end of inflation $\phi_e$ in \eq{phie}. Here we choose $\alpha =-1.2,-1.4,-1.6,-1.8,-2$, corresponding to the blue, red, green, cyan, and orange lines respectively. The other values chosen for the parameters are described at the end of Section~\ref{igcg}. Here it can be seen that the conditions for a burst of inflation, $a$ increasing and $|\eh|<1$, are satisfied for \aintthree. It can also be seen that $|\etah|>1$ and $|\xh|>1$ for \aconstraint~and $\alpha \le -3/2$, respectively.
  }
\label{fig:sr}
\end{figure}

\begin{figure}
	\centering
	\includegraphics[scale=1.]{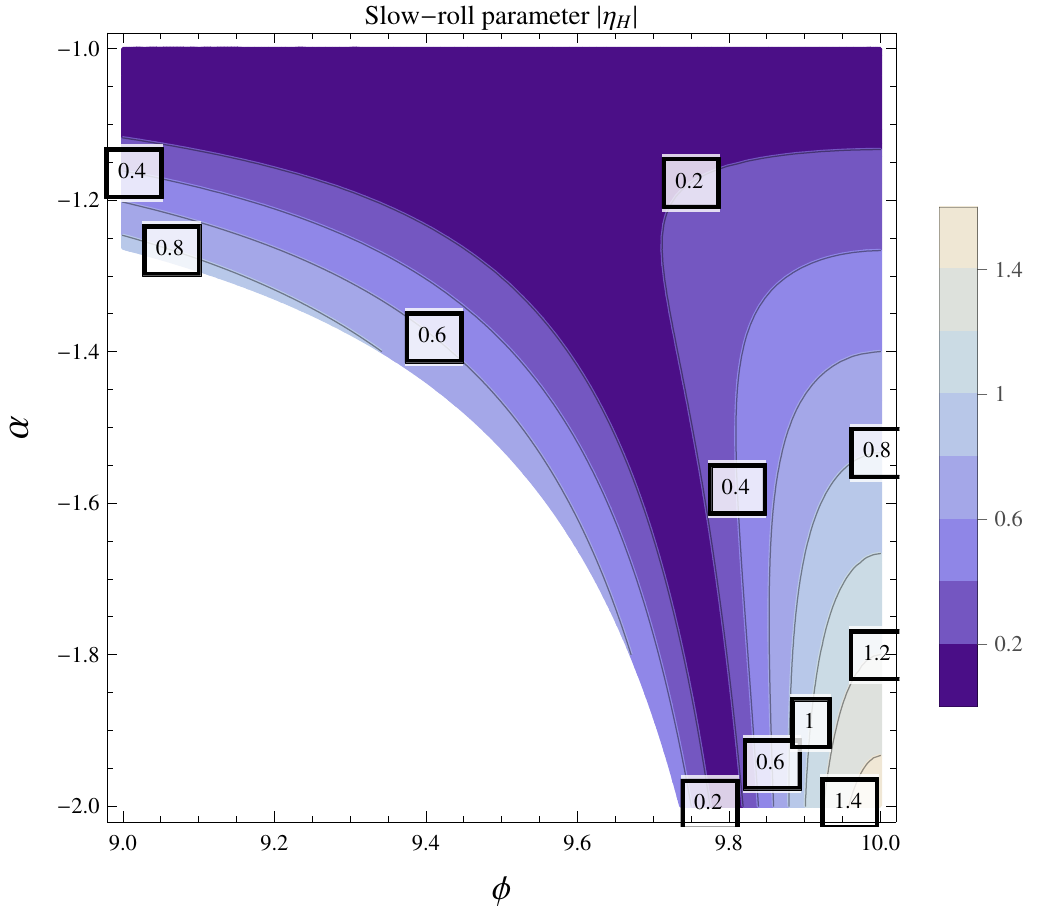}
	% epsilonHcontour.pdf: 0x0 pixel, 300dpi, 0.00x0.00 cm, bb=
	\caption{The contour values of the slow-roll parameter $\etah$ are plotted as functions of $\phi \le \phi_e$ (see \eq{phie}) and $-2<\alpha<-1$. Here it can be seen that $|\eta_H| >1$ for \aconstraint~in agreement with our estimation in \eq{etacond}.}
	\label{etacontour}
\end{figure}

\begin{figure}
	\centering
	\includegraphics[scale=1.]{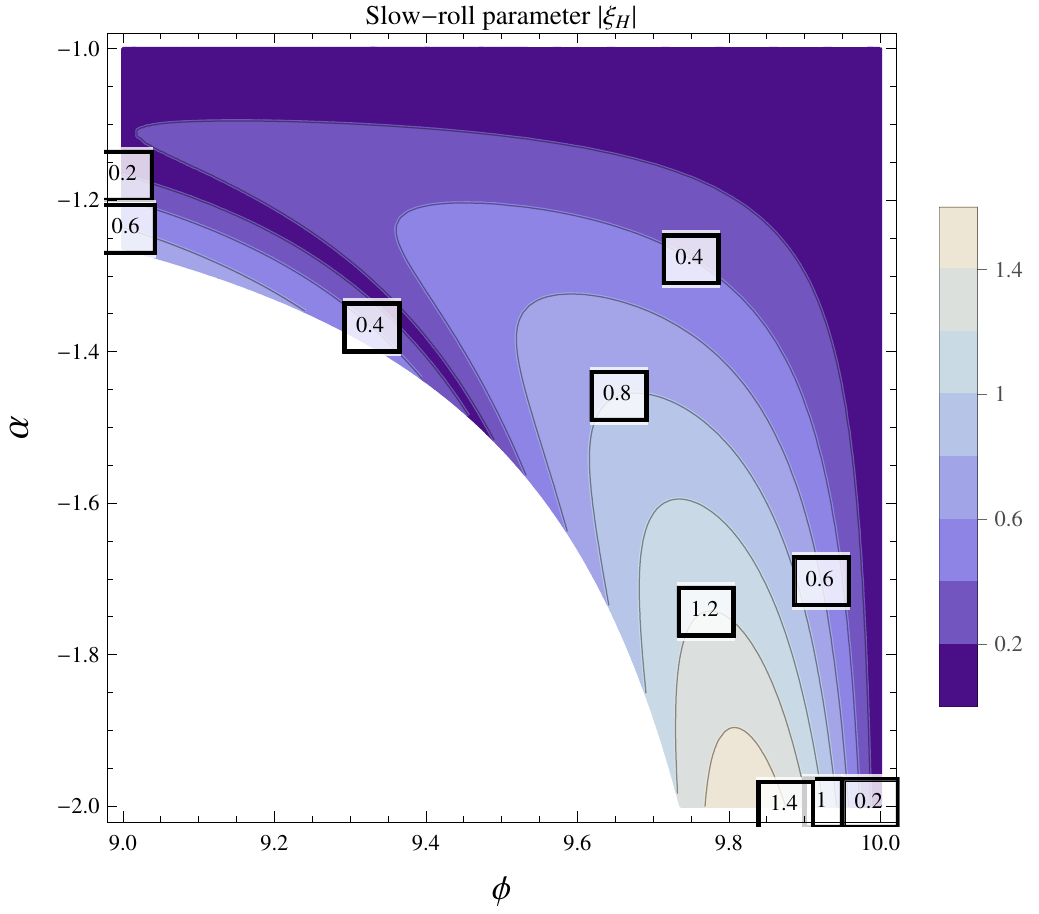}
	% epsilonHcontour.pdf: 0x0 pixel, 300dpi, 0.00x0.00 cm, bb=
	\caption{The contour values of the slow-roll parameter $|\xh|$ are plotted as functions of $\phi \le \phi_e$ (see \eq{phie}) and $-2<\alpha<-1$. Notice that $|\xh|$ is lager than unity for \aconstraintxi~in relative agreement with our estimation in \eq{xicond}.}
	\label{xi2contour}
\end{figure}

\begin{figure}
 \begin{minipage}{.45\textwidth}
  \includegraphics[scale=0.7]{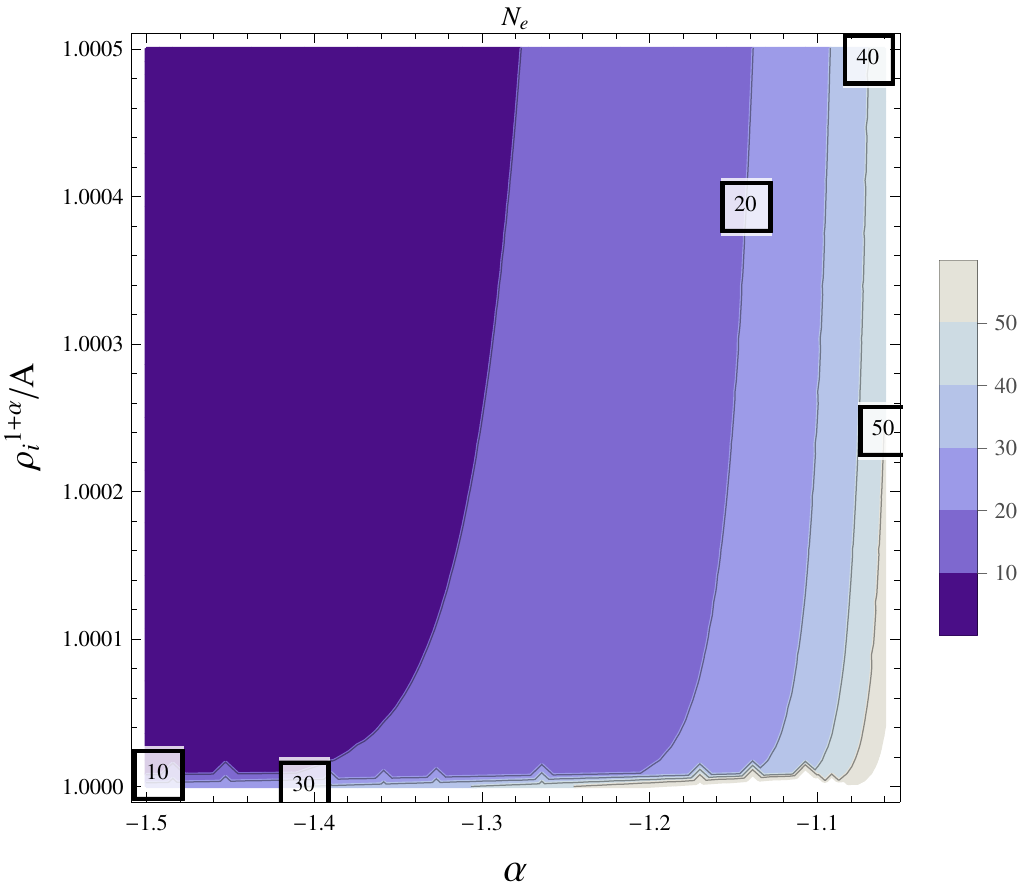}
  \end{minipage}
 \begin{minipage}{.45\textwidth}
  \includegraphics[scale=0.7]{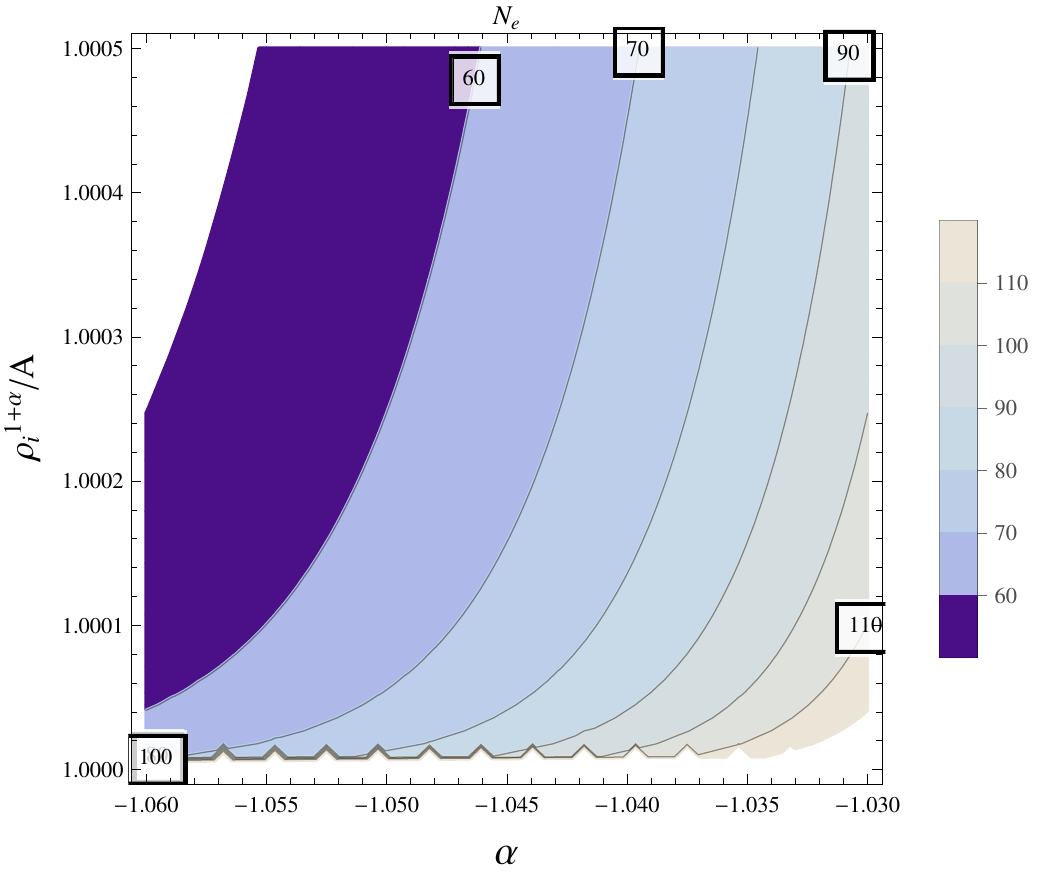}
 \end{minipage}
 \caption{The contour values of the number of $e$-folds at the end of inflation $N_e$ are plotted as functions of $\alpha$ and $\binvtext$ (equivalently $A$) for the GCG model using \eq{Ne}. Here it can be seen that even for $\alpha < -1.1$ and $\binvtext > 1.0001$ we have $N < 60$. The number of $e$-folds increases as either $\alpha \to -1$ or $\binvtext \to 1$, although the effect $\alpha \to -1$ is larger.}
\label{fig:MaxNe}
\end{figure}

\section{No slow-roll inflation à la Generalized Chaplygin Gas ($\alpha <-1$)} \label{nsrigcg}
In this section we first show that \theory~can indeed produce a burst of inflation when \aintthree. Then we prove that the slow-roll conditions $|\eta_H|<1$ and $|\xh|<1$ are only satisfied for the narrow range of values $-5/3 < \alpha < -1$ and $-3/2 < \alpha < -1$, respectively. Consequently we show that due to this behaviour of the slow-roll parameters we have that $N \ll 60$, thus the number of $e$-folds is not sufficient to solve the Big Bang problems during inflation unless $\alpha \to -1$. 
% Hence we conclude that the slow-roll approximations are not valid for the GCG model during inflation for most of the values of $\alpha$. 

In Fig. \ref{fig:sr} we plot the scale factor $a$ and the slow-roll parameters $\epsilon_H ,|\eta_H|$, and $|\xi_H|$ for several values of $\alpha$ and up to the end of inflation $\phi_e$ (see \eq{phie}). As it can be seen the conditions for a burst of inflation are satisfied since $a$ is increasing and $\epsilon_H <1$ for all values of \aintthree. There it can also be seen that $|\etah|>1$ and $|\xh|>1$ for \aconstraint~and $\alpha \le -3/2$, respectively. This behaviour of $\etah$ and its effect on the number of $e$-folds will be studied below.

Let us find the values of $\alpha$ for which \theory~fulfils $|\etah|<1$. In order to give an estimate, we consider $(1+\alpha)\Phi$ small. Then $\sech\left[(1+\alpha)\Phi\right] \sim 1$ such that $|\etah|<1$ in \eq{etach} becomes
\be 
|1+\alpha| <\frac{2}{3},
\ee
and, since we are only interested in \aintthree, we find that
\be\label{etacond}
|\etah|<1 \quad \mbox{only for} \quad -\frac{5}{3} < \alpha <-1.
\ee
Similarly the condition for $|\xh|<1$, and assuming $(1+\alpha)\Phi$ small and $\Phi \sim 1$ such that $\tanh^2\left[(1+\alpha)\Phi\right] \sim (1+\alpha)^2$, yields
\be 
\Bigl|(2-2\alpha-4\alpha^2)(1+\alpha)^2 \Bigr| < \frac{8}{9},
\ee
from which we have
\be\label{xicond}
|\xh|<1 \quad \mbox{only for} \quad -\frac{3}{2} < \alpha <-1.
\ee
In Figs. \ref{etacontour} and \ref{xi2contour} we plot respectively the contour values of $|\eta_H|$ and $|\xi_H|$ as functions of $\phi$ and $\alpha$. In the plots it can be seen that $|\eta_H| > 1$ and  $|\xh|>1$ for $\alpha<-5/3$ and $\alpha<-1.6 \approx -3/2$ , respectively, in agreement with our estimations. 

As anticipated, the behaviour of $\etah$ implies that the total number of $e$-folds $N$ would be too small for most of the parameter space of \theory. In Fig. \ref{fig:MaxNe} we plot the total number of $e$-folds at the end of inflation $N_e$ for the GCG model using \eq{Ne}. As it can be seen, in order to obtain at least $N_e > 60$ either $\alpha \to -1$ or $\binvtext \to 1$, although the effect $\alpha \to -1$ is larger. We recall that the case $\alpha=-1$ corresponds to a cosmological constant~\cite{Kamenshchik:2001cp, Bento:2002ps, Sen:2005sk, Dinda:2014zta, Chimento:2005au, 2009EPJC...63..349L, Fabris:2001tm, Makler:2002jv, Bilic:2001cg, Dev:2002qa}.

\subsection{Scalar and tensor perturbations}\label{stp}
We will now constrain the parameters of the GCG model from the study of the primordial perturbations. The evolution of the comoving curvature perturbation $\R$ is governed by the Mukhanov-Sasaki equation \cite{10.1143/PTP.70.394}
\be \label{mse} 
\frac{d^2 u_k}{d \tau^2} + \left(k^2 - \frac{1}{z}\frac{d^2z}{d\tau^2}\right) u_k=0, 
\ee
where $k$ is the comoving wave number, $u \equiv -{\text{sign}}(\dot \phi) z \R \mpl / \sqrt{8\pi}$ is the Mukhanov-Sasaki variable, and $z= a\sqrt{2\eh}$ such that
\begin{equation}
\label{zppz}
\frac{1}{z} \frac{d^2z}{d\tau^2} = 2a^2 H^2 \Bigl[ 1 + f_H(\phi(\tau)) \Bigr] \,,
\end{equation}
and where
\bea
f_H(\phi)  &&\equiv \epsilon_H -\frac{3}{2} \eta_H + \epsilon_H^2 -2 \epsilon_H \eta_H + \frac{1}{2} \eta_H^2 + \frac{1}{2} \xi_H^2 \,, \label{fH1}\\
&& =\frac{3}{8} \Bigl(3 (\alpha +1) (3 \alpha +2) \text{sech}^4 \big[(\alpha +1) \Phi \big]-\bigl(3 \alpha  (2 \alpha +5)+7 \bigr) \text{sech}^2\bigl[(\alpha +1) \Phi \big] -2 \Bigr) \label{fH2} ,
\eea
where we use the slow-roll parameters in \eqs{epsilonch} - \eqn{xich} in the last line. 
% The tensor perturbations $h$ follow a similar evolution to that of \eq{mse}~\cite{Akrami:2018vks, Akrami:2018odb}.
Despite its appearance as an expansion in the slow-roll parameters, \eq{zppz} is an exact expression. Here $f_H$ characterizes the validity of the slow-roll conditions, such that when $\{\eh, |\etah|, |\xh| \} <1$ we have $|f_H|\ll 1 $. 

\begin{figure}
 \includegraphics[scale=0.8]{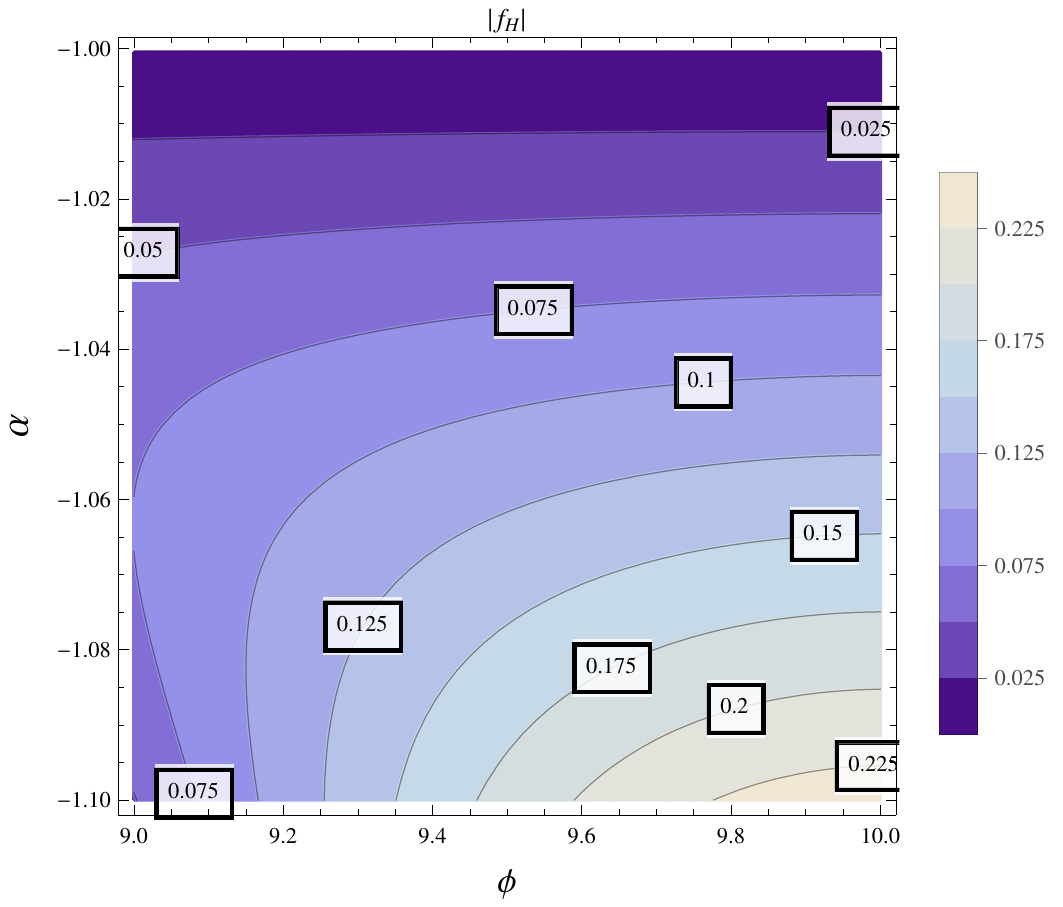}
 \caption{The contour values of $|f_H|$ in \eq{fH1} are plotted as functions of $\phi \le \phi_e (\alpha,A)$ and $-1.1<\alpha<-1.0005$. As can be seen only for $\alpha \gtrapprox -1.04$ \theory~can indeed have $|f_H| \ll 1$ such that $f_H$ can be neglected in \eq{zppz}. }
\label{fig:fcontour}
\end{figure}

When the slow-roll regime is satisfied, $f_H$ can be neglected and the solution to the Mukhanov-Sasaki equation can simply be written in terms of the Hankel function of the first kind $H_{\nu} (-k \tau)$. Nonetheless, as we have seen in the GCG model, the slow-roll approximations are only valid for $\alpha \to -1$ and $\binvtext \to 1$. For instance, following a similar procedure as before we can estimate that 
\be\label{fhcond}
|f_H| < 1 \quad \mbox{only for } \quad -\frac{\sqrt{17}}{3} < \alpha < -1,
\ee
which is an even tighter constrain than the one for the slow-roll parameters separately. In Fig. \ref{fig:fcontour} it can be seen that only for $\alpha \gtrapprox -1.04$ \theory~has $|f_H| \ll 1$ such that $f_H$ can be safely neglected in \eq{zppz}. 

For the rest of this section we will only consider $-1.04<\alpha <-1$ such that the slow-roll approximation is valid. Under this approximation, the spectrum of comoving curvature perturbations $\R$ and the spectrum of tensor perturbations $h$ are given by~\cite{Lidsey:1995np}
\bea
\label{pr}
P_\R (k_*)&=& \frac{1}{\pi \mpl^2} \frac{H^2}{\epsilon} \Biggl|_{k_* = a_* H_*} \, , \\
P_h (k_*)&=& \frac{16}{\pi \mpl^2}H^2\Bigl|_{k_* = a_* H_*} \,,
\eea
where the scalar and tensor power spectra are evaluated at the value of the inflaton field $\phi_*$ where the mode $k_* = a_* H_*$ crosses the Hubble radius for the first time~\cite{Akrami:2018vks, Akrami:2018odb}. 

Following Ref.~\cite{Akrami:2018odb} and using \eq{efoldsch}, we can relate the number of $e$-folds $N_{*}$ \textit{before the end} of inflation with the value of the field at horizon crossing $\phi_*$ by
\be \label{phis} 
\phi_* = \phi_0 + \sqrt{\frac{\mpl^2}{6\pi}} \frac{1}{1+\alpha} \arcsinh{\Biggl[ \sqrt{2 e^{3(1+\alpha)N_*}} \Biggr] } .
\ee 

The scalar spectral index $ n_{s}$ and the tensor-to-scalar ratio $r$ are given by~\cite{Lidsey:1995np}
\begin{eqnarray}
n_{s} - 1 &\equiv& \frac{d \ln P_\R}{d \ln k} = -4\epsilon_{H} +2\eta_{H} , \\
r &\equiv& \frac{P_h}{P_\R} = 16\epsilon_{H} ,
\end{eqnarray}
where again the quantities are evaluated at the horizon crossing scale $k_* = a_* H_*$. The Planck 2018 bounds on the power spectrum amplitude, spectral index, and tensor-to-scalar ratio are, respectively, $A_s=(2.0989 \pm 0.10141 ) \times 10^{-9}$,  $n_{s} = 0.9649 \pm 0.0042$, and $r< 0.064$\cite{Akrami:2018vks, Akrami:2018odb}. 
%%%%%%%%%%%%%%%%%%%%%%%%%%%%%%%%%%%%%%%%
\begin{figure}
\includegraphics[scale=0.6]{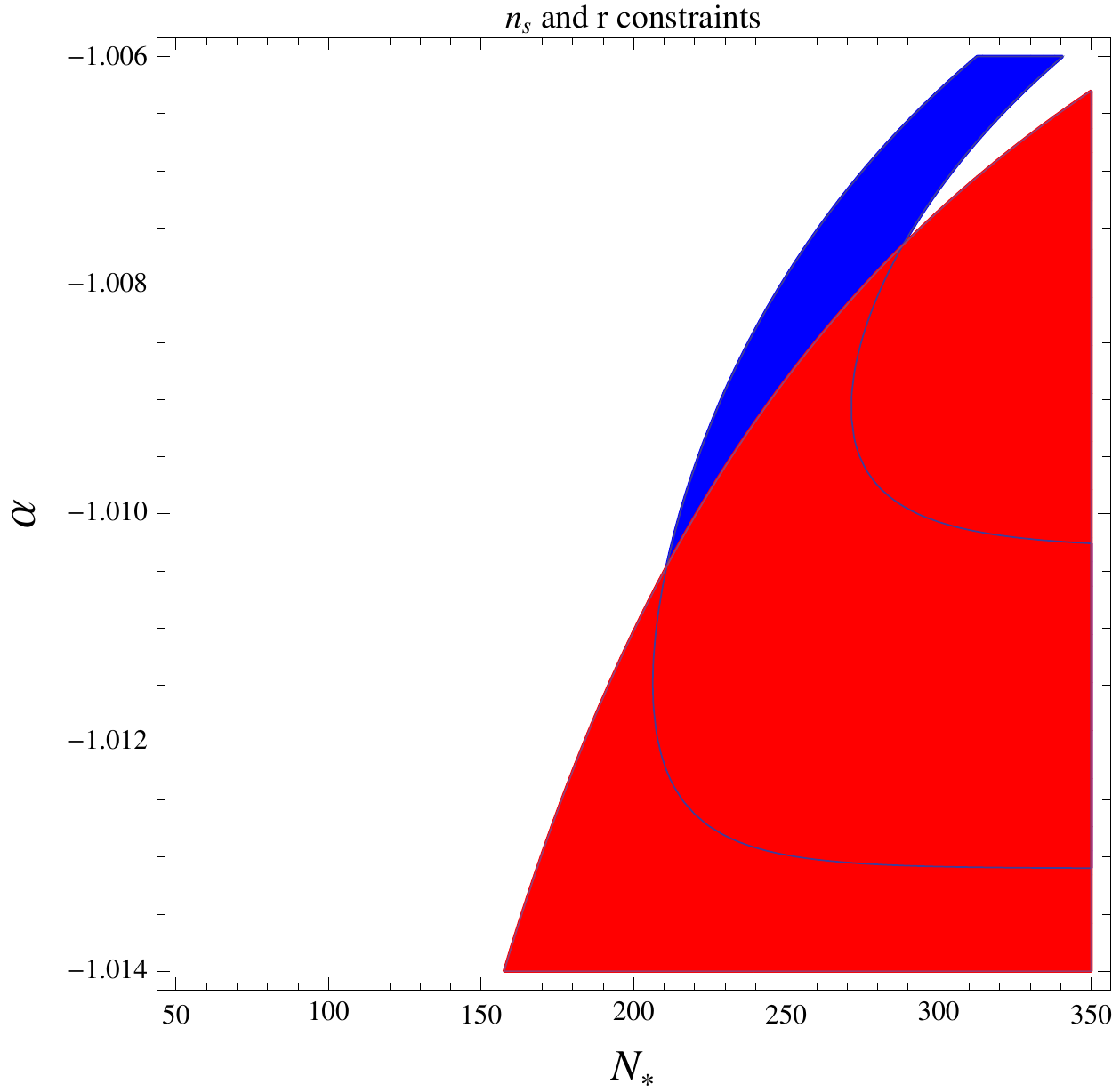}
\caption{The allowed region in the ($N_{*}$,$\alpha$) plane. The blue (red) region corresponds to the values of $N_*$ and $\alpha$ for which the constraints from Planck 2018 on $n_{s}$ ($r$) are fulfilled. Here it can be seen that, in order to simultaneously fulfill the constraints, the number of $e$-folds \textit{before the end} of inflation $N_*$ is way out the bounds $50<N_{*}<60$~\cite{Akrami:2018vks, Akrami:2018odb}.}
\label{fig:nsrregion}
\end{figure}
%%%%%%%%%%%%%%%%%%%%%%%%%%%%%%%%%%%%%%%%

Using the expression for $\phi_*$ in \eq{phis}, the scalar spectral index and tensor-to-scalar ratio at horizon crossing can be written, respectively, as
\begin{eqnarray}
\label{nsNs}
n_{s} (N_*,\alpha) &=& -2 + \frac{3(2+\alpha)}{1 +2 e^{ 3 (1+\alpha) N_{*}}} ,\\
\label{rNs}
r (N_*,\alpha)&=& \frac{24}{ 1 + \frac12 e^{-3 (1+\alpha) N_{*}} } .
\end{eqnarray}
In Fig.~\ref{fig:nsrregion} we plot the allowed region in the ($N_{*}$,$\alpha$) plane satisfying the above Planck 2018 constraints. Since the relevant scales that can be probed through CMB observations are those that exited the horizon around $50<N_{*}<60$~\cite{Akrami:2018vks, Akrami:2018odb}, it can be seen in Fig.~\ref{fig:nsrregion} that the GCG model is ruled out by the Planck 2018 constraints. Notice that these results are independent of the parameter $A$, thus the constraints on $A$ from the Planck 2018 bounds are not relevant.

In this section we have seen that the slow-roll condition $|\etah|<1$ is not satisfied by the GCG model when \aconstraint~and thus $N_e(\alpha, A) \ll 60$. In fact we have seen that only for $\alpha \to -1$ we have $N_e(\alpha, A)>60$. Moreover, from the Planck 2018 results we found that the model is ruled out. Our results differ from those in Ref.~\cite{Dinda:2014zta} since in there the authors used the BICEP2 constraint of $r= 0.2_{-0.05}^{+0.07}$.

\section{No slow-roll inflation à la Generalized Chaplygin-Jacobi Gas} \label{nsrigcjggr}
In this section we study the GCJG model in \eq{eoschj} and extend the analysis of Refs.~\cite{Villanueva:2015ypa, Villanueva:2015xpa} from \ainttwo~to \anewint. Here we show that, even when the extra parameter $\kappa$ is added, \theoryj~does not produce inflation for \aint~and that the slow-roll conditions for $|\eta_H|$ and $|\xh|$ are only satisfied in a narrow range of values of $\alpha$ very close to -1. %Similar to the case of the GCG model, we will see that since $|\eta_H| > 1$ the total number of $e$-folds after inflation $N_e$ will be small. 

\begin{figure}
 \begin{minipage}{.45\textwidth}
  \includegraphics[scale=0.56]{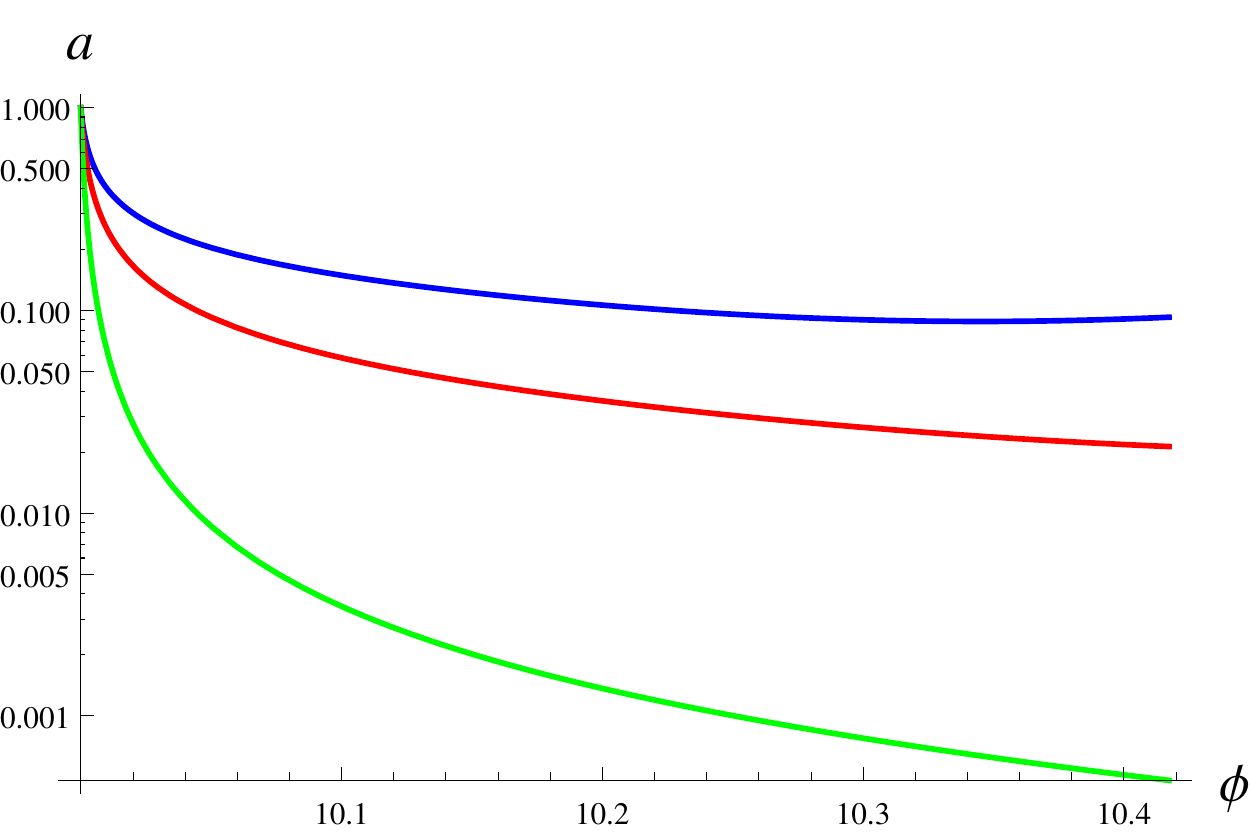}	
  \end{minipage}
 \begin{minipage}{.45\textwidth}
  \includegraphics[scale=0.56]{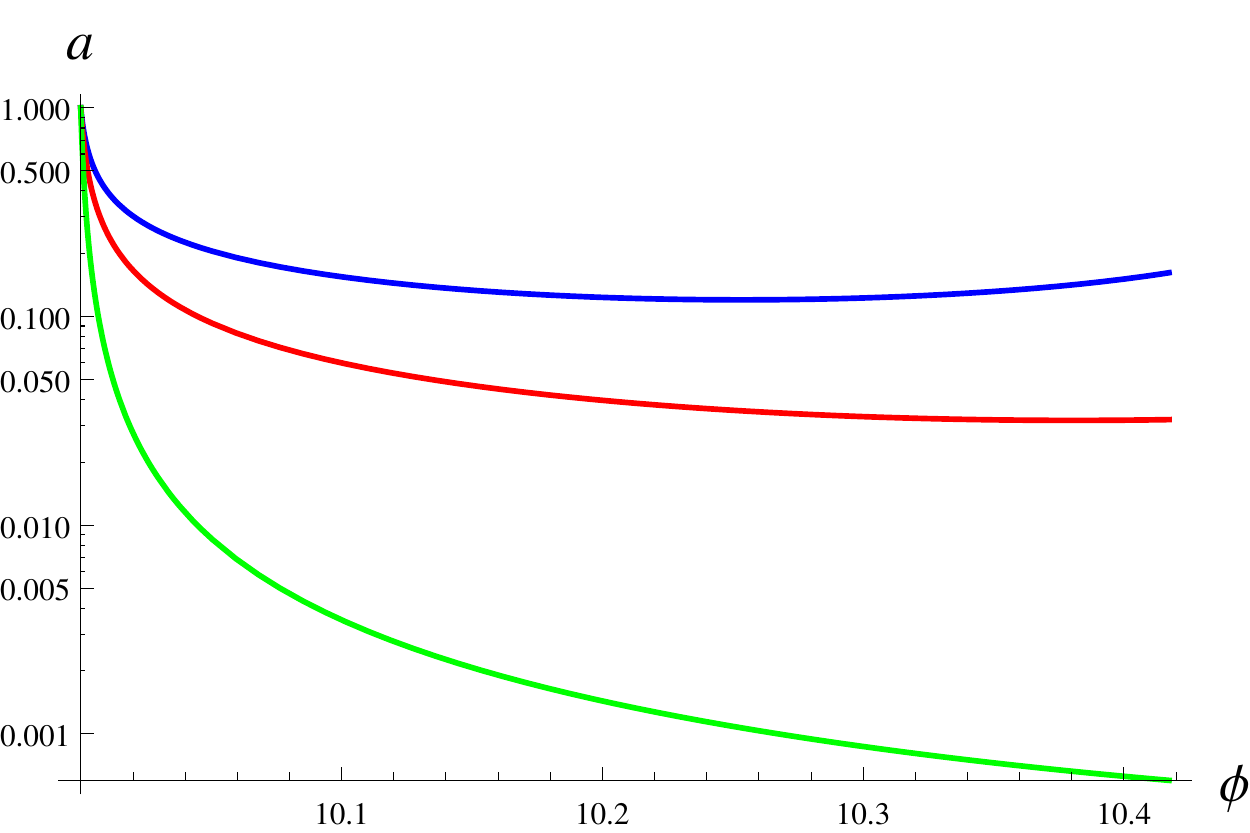}
 \end{minipage}
  \caption{The scale factor $a$ is plotted for $\alpha=0.5, 0, -0.5$ and $\binvtext=1.00001$ in the blue, red, and green lines, respectively. On the left (right) panel we use $\kappa = 0.8 (0.2)$. It can be seen that the scale factor is a decreasing function of $\phi$, hence there is no inflationary period for $\alpha>-1$.}
\label{fig:anij}
\end{figure}

In the GCJG the generating function in \eq{Hch} is written as \cite{Villanueva:2015ypa, Villanueva:2015xpa}
\begin{equation}
\label{Hchj}
H(\phi, \kappa)=H_0\,\textrm{nc}^{\frac{1}{1+\alpha}}\left[(1+\alpha)\,\Phi\right],
\end{equation}
where $\textrm{nc}(x)=1/\textrm{cn}(x)$, and $\textrm{cn}(x)\equiv\textrm{cn}(x|\, \kappa)$ is the Jacobi elliptic cosine function, and $\kappa$ is the modulus. The generating function of the GCG in \eq{Hch} is recovered when $\kappa=1$. Similarly as in \theory~case the scale factor is now
\be \label{achj}
a(\phi) = a_i \Biggl( \frac{\sd[(1+ \alpha) \Phi]}{\sd[(1+ \alpha) \Phi_i]} \Biggr)^{\frac{-2}{3(1+\alpha)}},
\ee
where $\textrm{sd}(x)\equiv \textrm{sd}(x|\,\kappa)=\textrm{sn}(x)/\textrm{dn}(x)$. Notice the minus sign in the exponent, which is also missing in Eq.~(3.14) of \refc{Villanueva:2015ypa}. In Fig.~\ref{fig:anij} we plot the scale factor as a function of $\phi$ which shows that $a$ is indeed decreasing for \aint. 

The slow-roll parameters are now given by
\bea
\label{epsilonchj}
\epsilon_H &=&
\frac{3}{2}\frac{\text{dn}^{2}\left[(1+\alpha)\,\Phi\right]\,
\text{sn}^{2}\left[(1+\alpha)\,\Phi\right]}{\text{cn}^{2}\left[(1+\alpha)\,\Phi\right]} , \\ 
\label{etachj}
\eta_H &=& \epsilon_H\left\{ \frac{1+\alpha \,\textrm{cn}^2[(1+\alpha)\Phi]}{\textrm{sn}^2[(1+\alpha) \Phi]}+\frac{(1+\alpha)\,(1-\kappa)}{\textrm{dn}^2[(1+\alpha) \Phi]}\right\}, \\
\label{xichj}
\xi_H^2 &=& \epsilon_H^2\left\{
\frac{(2\alpha^2+7\alpha+6)\,(1-\kappa)}{\textrm{sn}^2\,[(1+\alpha)\Phi]\,\textrm{dn}^2\,[(1+\alpha)\Phi]}+\frac{3}{2\epsilon_H}
\Bigl( (2 \kappa-1)-\kappa \,\alpha\,(1+2\alpha)\,\textrm{cn}^2\,[(1+\alpha)\Phi] \Bigr)
\right\},
\eea
where $\textrm{sn}(x)\equiv \textrm{sn}(x|\,\kappa)$ and $\textrm{dn}(x)\equiv \textrm{dn}(x|\,\kappa)$ are the Jacobi elliptic sine and delta functions, respectively. The condition for the end of inflation, $\epsilon_{_{H}}(\Phi_e) = 1$, now yields \cite{Villanueva:2015ypa}
\be \label{phiechj}
\Phi_e (\alpha, \kappa)=\frac{1}{1+\alpha} F\left[\arcsin \left( \sqrt{ y }\right),\,\kappa\right],
\ee
where $F(\varphi,\,\kappa)$ is the normal elliptic integral of the first kind and $y=[5-(25-24 \kappa)^{1/2}]/(6 \kappa)$.

\begin{figure}
	\centering
	\begin{minipage}{.45\textwidth}
	\includegraphics[scale=0.7]{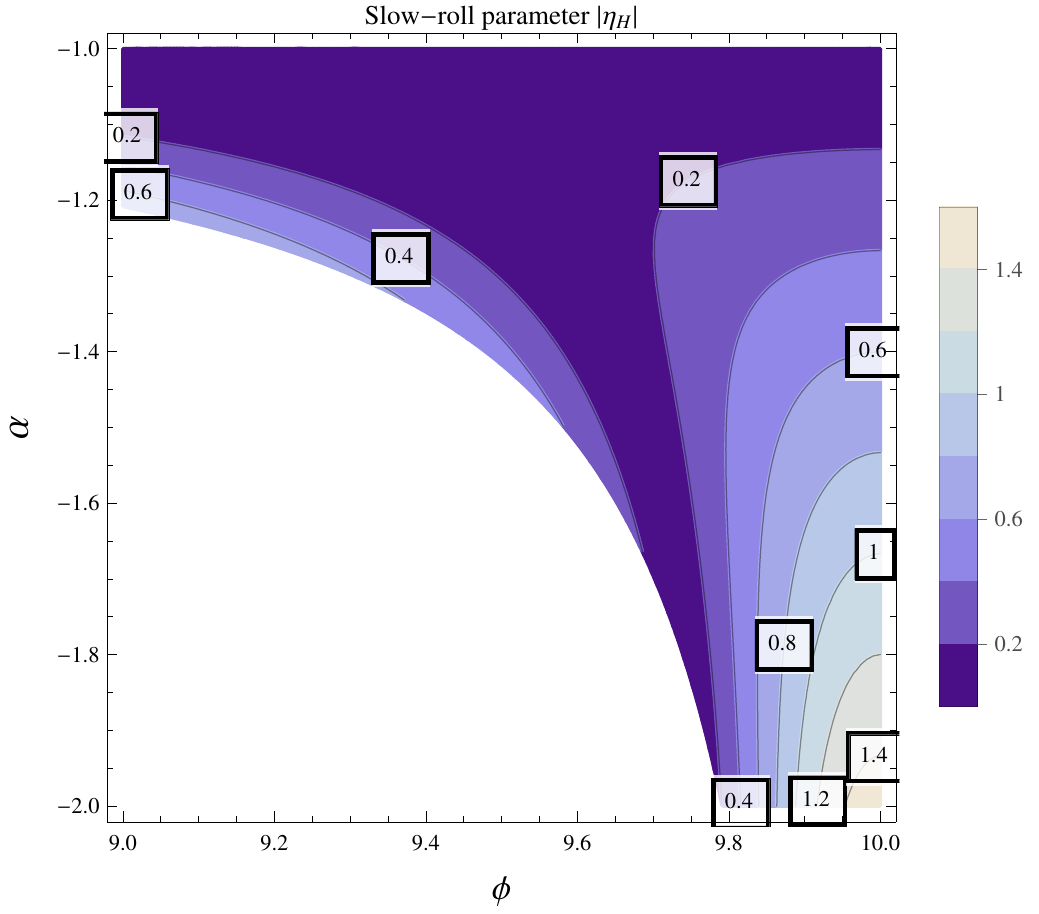}
	\end{minipage}
	\begin{minipage}{.45\textwidth}
	\includegraphics[scale=0.7]{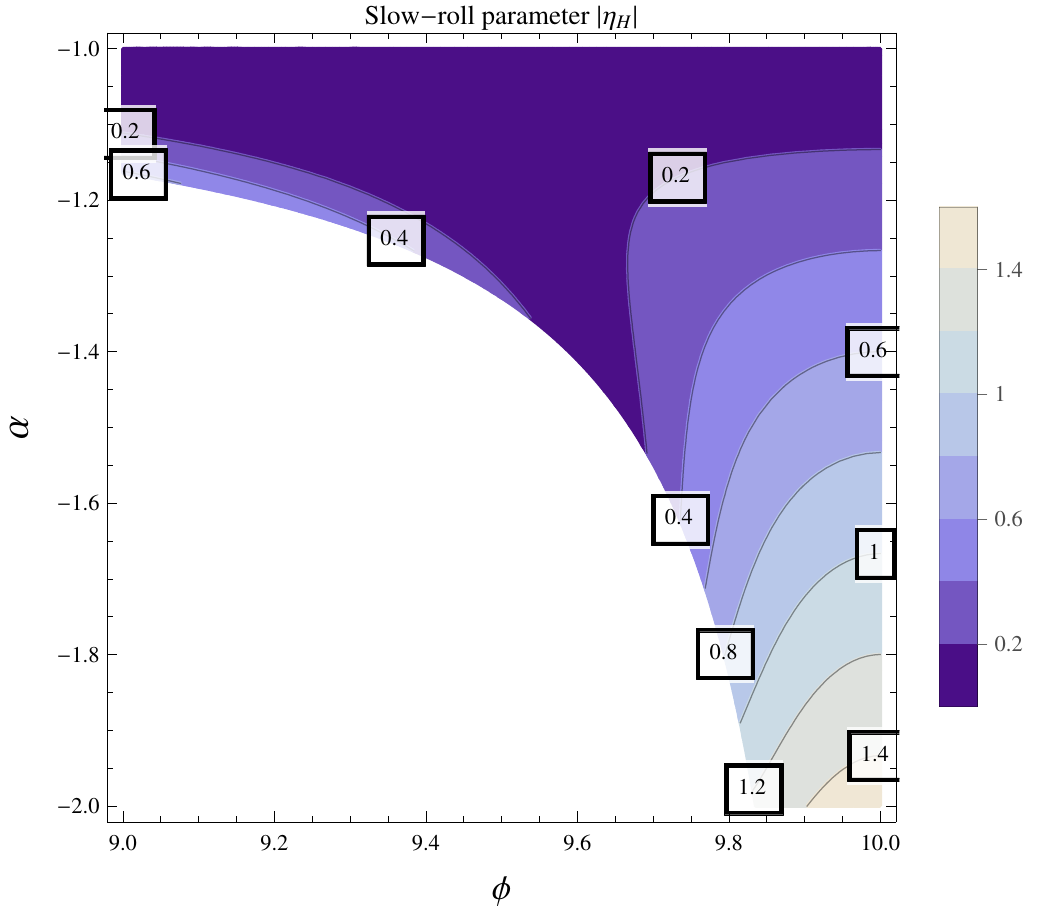}
	\end{minipage}
	% epsilonHcontour.pdf: 0x0 pixel, 300dpi, 0.00x0.00 cm, bb=
	\caption{The contour values of the slow-roll parameter $\etah$ are plotted as functions of $ \phi  \le \phi_e (\alpha,\kappa)$ and $-2<\alpha<-1$. On the left (right) we use $\kappa= 0.8 (0.2)$. Here it can be seen that $|\eta_H| >1$ for \aconstraint.}
	\label{fig:etacontourchj}
\end{figure}

The number of $e$-folds is now given by \cite{Villanueva:2015ypa}
\begin{equation}
\label{efoldschj}
N(\phi)=-\frac{2}{3(1+\alpha)}\ln\left\{\frac{\textrm{sd}[(1+\alpha)\Phi]}{\textrm{sd}[(1+\alpha)\Phi_i]}\right\}.
\end{equation}
Then the total number of $e$-folds at the end of inflation is
\be \label{Nej} 
N_e(\alpha, A, \kappa) = \frac{1}{3 (1+\alpha)}\,\ln\left\{\frac{ 1 }{\fn(\kappa)} \Biggl(\binv - 1 \Biggr) \right\},
\ee
where $\fn\equiv\sd^2 \Bigl[  F\left[\arcsin \left( \sqrt{ y(\kappa) }\right),\, \kappa \right] \Bigr]$. This last equation should be compared with \eq{Ne} where $\fn$ plays the role of the factor 2 in the denominator. In fact in the limits $\kappa \to 1$ and $\kappa \to 0$, we have $\fn \to 2$ and $\fn\to 0.4$ respectively, thus that $\binvtext$ is now reduced to the boundary $1< \binvtext <1.4$. This means that the introduction of the new parameter $\kappa$ does not solve the previous problems of $N\ll60$ for $\alpha < -5/3$ since  $N_e$ does not strongly depend on $\kappa$.

Let us now find an approximation for the values of $\alpha$ such that $|\eta_H|< 1$. Similar to Section~\ref{nsrigcg}, we expand $\etah$ in \eq{etachj} for $(1+\alpha)\Phi$ small and obtain a similar condition to that in \eq{etacond}, i.e.
\be
|\etah|<1 \quad \mbox{only for} \quad -\frac{5}{3} < \alpha <-1.
\ee
In Fig. \ref{fig:etacontourchj} we plot the contour values of $\eta_H$ as function of $\phi$ and $\alpha$. There it can be seen that $|\eta_H|> 1$ for all \aconstraint. As we stated before, the introduction of the new parameter $\kappa$ does not resolve the slow-roll problem for $\alpha < -5/3$. In fact, similar results to the ones in Sec.~\ref{nsrigcg} are found for $\xi_H$, $N_e$, $f_H$, and the parameters of the model.

\section{Conclusions} \label{conclusions}
In this paper we have studied the GCG and GCJG models for primordial inflation in the context of GR. We have shown that the models do not produce inflation when $-1<\alpha<1$, contrary to the results in Refs.~\cite{delCampo:2013uta, Villanueva:2015ypa, Villanueva:2015xpa}. Then we show that,  although there is a burst of inflation for \aintthree, the slow-roll parameters $\etah$ and $\xh$ are larger than unity for $\alpha < -5/3$ and $\alpha <-8/5$, respectively. Furthermore, they are much less than unity throughout inflation only for $-1.1<\alpha <-1$. Hence the parameter space of the GCG and GCJG models is greatly reduced. 

Then we showed that such a behaviour of the slow-roll parameters implies that the total number of $e$-folds $N$ during inflation is $N \ll 60$ for most of their parameter space, especially for $\alpha < -1.1$. Such that the horizon and flatness problems of the Big Bang cosmology cannot be solved during inflation by the GCG and GCJG models when $\alpha < -1.1$, since the theoretical bound on the number of $e$-folds is $50 < N_* < 60$ \cite{Akrami:2018vks, Akrami:2018odb}. 

We also constrained the parameters of the models from the Planck 2018 results. From the bounds on the scalar spectral index $n_s$ and the tensor-to-scalar ratio $r$, we found that the models are ruled out since they require more than $200~e$-folds before the end of inflation in order to fulfill the bounds.

We conclude that the violation of the slow-roll conditions is a generic feature of the models during inflation when GR is considered and that they are ruled out by the Planck 2018 results.

\section{Acknowledgements}
We thank the anonymous referee for valuable remarks and comments which significantly contributed to improve the paper. We also thank Ramón Herrera for useful discussions and Clément Stahl and Gabriel Gómez for useful comments on an early version of the manuscript. A.G.C. was supported by Beca de Inicio Postdoctoral REXE RA N° 315/5398/2019 Universidad de Valparaíso.

\bibliography{Bibliography,BibliographyGCG}

\begin{thebibliography}{10}

\bibitem{Akrami:2018vks}
Planck, Y.~Akrami {\em et~al.},
\newblock (2018), arXiv:1807.06205.
%%CITATION = ARXIV:1807.06205;%%

\bibitem{Akrami:2018odb}
Planck, Y.~Akrami {\em et~al.},
\newblock (2018), arXiv:1807.06211.
%%CITATION = ARXIV:1807.06211;%%

\bibitem{Guth:1980zm}
A.~H. Guth,
\newblock Phys.Rev. {\bf D23}, 347 (1981).
%%CITATION = PHRVA,D23,347;%%

\bibitem{Linde:1981mu}
A.~D. Linde,
\newblock Phys.Lett. {\bf B108}, 389 (1982).
%%CITATION = PHLTA,B108,389;%%

\bibitem{Starobinsky:1998mj}
A.~A. Starobinsky,
\newblock Grav.Cosmol. {\bf 4}, 88 (1998), arXiv:astro-ph/9811360.
%%CITATION = ASTRO-PH/9811360;%%

\bibitem{Martin:2013tda}
{Martin Jerome et al.},
\newblock Phys. Dark Univ. {\bf 5-6}, 75 (2014), arXiv:1303.3787.
%%CITATION = ARXIV:1303.3787;%%

\bibitem{Wang:2013eqj}
Y.~Wang,
\newblock Commun. Theor. Phys. {\bf 62}, 109 (2014), arXiv:1303.1523.
%%CITATION = ARXIV:1303.1523;%%

\bibitem{Castiblanco:2019mgb}
L.~Castiblanco, R.~Gannouji, and C.~Stahl,
\newblock {Large scale structures: from inflation to today: a brief report},
\newblock 2019, arXiv:1910.03931.
%%CITATION = ARXIV:1910.03931;%%

\bibitem{xc}
X.~Chen,
\newblock Adv.Astron. {\bf 2010}, 638979 (2010), arXiv:1002.1416.
%%CITATION = ARXIV:1002.1416;%%

\bibitem{Chakraborty:2018scm}
S.~Chakraborty, T.~Paul, and S.~SenGupta,
\newblock Phys. Rev. {\bf D98}, 083539 (2018), arXiv:1804.03004.
%%CITATION = ARXIV:1804.03004;%%

\bibitem{Bento:2002ps}
M.~C. Bento, O.~Bertolami, and A.~A. Sen,
\newblock Phys. Rev. {\bf D66}, 043507 (2002), arXiv:gr-qc/0202064.
%%CITATION = GR-QC/0202064;%%

\bibitem{Sen:2005sk}
A.~A. Sen and R.~J. Scherrer,
\newblock Phys. Rev. {\bf D72}, 063511 (2005), arXiv:astro-ph/0507717.
%%CITATION = ASTRO-PH/0507717;%%

\bibitem{Dinda:2014zta}
B.~R. Dinda, S.~Kumar, and A.~A. Sen,
\newblock Phys. Rev. {\bf D90}, 083515 (2014), arXiv:1404.3683.
%%CITATION = ARXIV:1404.3683;%%

\bibitem{Chimento:2005au}
L.~P. Chimento and R.~Lazkoz,
\newblock Class. Quant. Grav. {\bf 23}, 3195 (2006), arXiv:astro-ph/0505254.

\bibitem{Kamenshchik:2001cp}
A.~{\relax Yu}. Kamenshchik, U.~Moschella, and V.~Pasquier,
\newblock Phys. Lett. {\bf B511}, 265 (2001), arXiv:gr-qc/0103004.
%%CITATION = GR-QC/0103004;%%

\bibitem{2009EPJC...63..349L}
J.~{Lu}, Y.~{Gui}, and L.~X. {Xu},
\newblock European Physical Journal C {\bf 63}, 349 (2009), arXiv:1004.3365.

\bibitem{Fabris:2001tm}
J.~C. Fabris, S.~V.~B. Goncalves, and P.~E. de~Souza,
\newblock Gen. Rel. Grav. {\bf 34}, 53 (2002), arXiv:gr-qc/0103083.
%%CITATION = GR-QC/0103083;%%

\bibitem{Makler:2002jv}
M.~Makler, S.~Quinet~de Oliveira, and I.~Waga,
\newblock Phys. Lett. {\bf B555}, 1 (2003), arXiv:astro-ph/0209486.
%%CITATION = ASTRO-PH/0209486;%%

\bibitem{Bilic:2001cg}
N.~Bilic, G.~B. Tupper, and R.~D. Viollier,
\newblock Phys. Lett. {\bf B535}, 17 (2002), arXiv:astro-ph/0111325.
%%CITATION = ASTRO-PH/0111325;%%

\bibitem{Dev:2002qa}
A.~Dev, D.~Jain, and J.~S. Alcaniz,
\newblock Phys. Rev. {\bf D67}, 023515 (2003), arXiv:astro-ph/0209379.
%%CITATION = ASTRO-PH/0209379;%%

\bibitem{Herrera:2008us}
R.~Herrera,
\newblock Phys. Lett. {\bf B664}, 149 (2008), arXiv:0805.1005.
%%CITATION = ARXIV:0805.1005;%%

\bibitem{Mak:2005iq}
M.~K. Mak and T.~Harko,
\newblock Phys. Rev. {\bf D71}, 104022 (2005), arXiv:gr-qc/0505034.
%%CITATION = GR-QC/0505034;%%

\bibitem{Jawad:2016key}
A.~Jawad, A.~Ilyas, and S.~Rani,
\newblock Int. J. Mod. Phys. {\bf D26}, 1750031 (2016), arXiv:1603.08798.
%%CITATION = ARXIV:1603.08798;%%

\bibitem{Barreiro:2004bd}
T.~Barreiro and A.~A. Sen,
\newblock Phys. Rev. {\bf D70}, 124013 (2004), arXiv:astro-ph/0408185.
%%CITATION = ASTRO-PH/0408185;%%

\bibitem{Bertolami:2006zg}
O.~Bertolami and V.~Duvvuri,
\newblock Phys. Lett. {\bf B640}, 121 (2006), arXiv:astro-ph/0603366.
%%CITATION = ASTRO-PH/0603366;%%

\bibitem{delCampo:2013uta}
S.~del Campo,
\newblock JCAP {\bf 1311}, 004 (2013), arXiv:1310.4988.
%%CITATION = ARXIV:1310.4988;%%

\bibitem{Villanueva:2015ypa}
J.~R. Villanueva,
\newblock JCAP {\bf 1507}, 045 (2015), arXiv:1505.03107.
%%CITATION = ARXIV:1505.03107;%%

\bibitem{Villanueva:2015xpa}
J.~R. Villanueva and E.~Gallo,
\newblock Eur. Phys. J. {\bf C75}, 256 (2015), arXiv:1505.03096.
%%CITATION = ARXIV:1505.03096;%%

\bibitem{Starobinsky:1992ts}
A.~A. Starobinsky,
\newblock JETP Lett. {\bf 55}, 489 (1992).
%%CITATION = JTPLA,55,489;%%

\bibitem{Adams:2001vc}
J.~A. Adams, B.~Cresswell, and R.~Easther,
\newblock Phys. Rev. {\bf D64}, 123514 (2001), arXiv:astro-ph/0102236.
%%CITATION = ASTRO-PH/0102236;%%

\bibitem{Lidsey:1995np}
J.~E. Lidsey {\em et~al.},
\newblock Rev.Mod.Phys. {\bf 69}, 373 (1997), arXiv:astro-ph/9508078.
%%CITATION = ASTRO-PH/9508078;%%

\bibitem{inf}
A.~R. Liddle and D.~H. Lyth,
\newblock {\em {Cosmological inflation and large scale structure}} (, 2000).
%%CITATION = ISBN-13-9780521828499;%%

\bibitem{Kinney:1997ne}
W.~H. Kinney,
\newblock Phys. Rev. {\bf D56}, 2002 (1997), arXiv:hep-ph/9702427.
%%CITATION = HEP-PH/9702427;%%

\bibitem{Romano:2008rr}
A.~E. Romano and M.~Sasaki,
\newblock Phys.Rev. {\bf D78}, 103522 (2008), arXiv:0809.5142.
%%CITATION = ARXIV:0809.5142;%%

\bibitem{Arroja:2011yu}
F.~Arroja, A.~E. Romano, and M.~Sasaki,
\newblock Phys.Rev. {\bf D84}, 123503 (2011), arXiv:1106.5384.
%%CITATION = ARXIV:1106.5384;%%

\bibitem{a1}
P.~Adshead, C.~Dvorkin, W.~Hu, and E.~A. Lim,
\newblock Phys.Rev. {\bf D85}, 023531 (2012), arXiv:1110.3050.
%%CITATION = ARXIV:1110.3050;%%

\bibitem{a2}
X.~Chen, R.~Easther, and E.~A. Lim,
\newblock JCAP {\bf 0706}, 023 (2007), arXiv:astro-ph/0611645.
%%CITATION = ASTRO-PH/0611645;%%

\bibitem{a3}
X.~Chen, R.~Easther, and E.~A. Lim,
\newblock JCAP {\bf 0804}, 010 (2008), arXiv:0801.3295.
%%CITATION = ARXIV:0801.3295;%%

\bibitem{Romano:2014kla}
A.~G. Cadavid and A.~E. Romano,
\newblock Eur. Phys. J. {\bf C75}, 589 (2015), arXiv:1404.2985.
%%CITATION = ARXIV:1404.2985;%%

\bibitem{GallegoCadavid:2017pol}
A.~Gallego~Cadavid,
\newblock J. Phys. Conf. Ser. {\bf 831}, 012003 (2017), arXiv:1703.04375.
%%CITATION = ARXIV:1703.04375;%%

\bibitem{10.1143/PTP.70.394}
M.~Sasaki,
\newblock Progress of Theoretical Physics {\bf 70}, 394 (1983).

\end{thebibliography}
\bibliographystyle{h-physrev4}
\end{document}